      [1999/12/01 v1.4c Il Nuovo Cimento]

\documentclass{cimento}

\usepackage{graphicx,amssymb}

\newcommand{\apj}{Astrophys. J.} 	    
\newcommand{\aap}{Astron. and Astrophys.} 	    
\newcommand{\mnras}{MNRAS} 	    
\newcommand{\degrees}{\mbox{$^{\circ}$}}    
\newcommand{\swift}{{\it Swift}}

\title{Swift X-ray Afterglows: Where are the X-ray Jet Breaks?}
\shorttitle{Where are the X-ray Jet Breaks?}
\author{D.~N.~Burrows and J. Racusin
}
\instlist{\inst
  Department of Astronomy \& Astrophysics, The
  Pennsylvania State University, 525 Davey Lab, University Park, PA
  16802  USA
  }
\PACSes{\PACSit{98.70.Rz}{$\gamma$-ray sources; $\gamma$-ray bursts}
\PACSit{95.85.Nv}{X-ray observations}}

\begin{document}

\maketitle

\begin{abstract}
We examine the {\it Swift}/X-ray Telescope (XRT) light curves from the
first $\sim 150$ gamma-ray burst (GRB) afterglows.  Although we expected to find jet breaks at
typical times of 1-2 days after the GRB, we find that these appear to
be extremely rare.  Typical light curves have a break in the slope at
about $10^4$~s, followed by a single power-law decay whose slope is
much too shallow to be consistent with expectations for jet breaks.
X-ray light curves typically extend out to $\sim 10$~days without any
further breaks, until they become too faint for the XRT to detect.  In
some extreme cases, light curves extend out to more than two months
without evidence for jet breaks.  This raises concerns about
our understanding of afterglow and jet dynamics, and of GRB energetics.
\end{abstract}

\section{Introduction}
The beaming factors of Gamma-Ray Bursts (GRBs) are critically important
for understanding their overall energetics.  For GRBs at known
redshift, the GRB fluence can be converted to a total radiated energy
assuming isotropic radiation.  The values found for $E_{\gamma,iso}$
can range up to $10^{54}$ ergs, a value that is difficult to
explain unless the radiation is actually concentrated into a narrow
beam, or jet, pointed towards us.  The jet beaming factor must then be
measured in order to determine the actual beamed energy radiated by
the GRB.

Fortunately, the jet opening angle can be determined directly from
detailed measurements of the light curves of GRBs.
The power law decay indices of GRB afterglows are
expected to steepen to $F_\nu \propto t^{-p}$ when the jet has
decelerated to the point that the bulk Lorentz factor, $\Gamma$, is
given by $\Gamma \sim \theta_j^{-1}$, where $p$ is the power
law index of the electron energy distribution and $\theta_j$ is the
opening angle of the collimated jet \cite{Sari99}.
The jet opening angle for a uniform jet can be expressed as
\begin{equation}
\theta_j = \frac{1}{6} \left( \frac{t_j}{1+z}\right)^{3/8}\left( \frac{n \eta_\gamma}{E_{\gamma,iso,52}}\right)^{1/8},
\end{equation}
where $\theta_j$ is the jet opening angle in radians, $t_j$ is the jet
break time in days in the observer's frame, $n$ is the ambient
number density in cm$^{-3}$, $\eta_\gamma$ is the radiation efficiency, and
$E_{\gamma,iso,52}$ is the isotropic equivalent energy radiated in
gamma rays in
units of $10^{52}$~erg \cite{Sari99, Frail01}.
Without broad-band data, including radio observations,
fitted to a detailed afterglow model, we cannot in general determine the
density or efficiency for a given burst, and only some 10\% of bursts
have radio detections.  Furthermore, the redshift is 
known for only $\frac{1}{3}$ of bursts with XRT observations.
We therefore rewrite Eq. 1 in terms of typical values for the unknown
parameters, as
\begin{equation}
\theta_j = 0.064~~\xi~t_j^{3/8}, 
\end{equation}
where we define
\begin{equation}
\xi \equiv \left( \frac{3.5}{1+z}\right)^{3/8}\left(
  \frac{\eta_\gamma}{0.2}\right)^{1/8} \left( \frac{n}{E_{\gamma,iso,53}}\right)^{1/8}.
\end{equation}
For typical values of $z \sim 2.5$ (the mean redshift for long Swift bursts),
$\eta_\gamma \sim 0.2$, $n \sim 1$~cm$^{-3}$, and $E_{\gamma,iso} \sim
10^{53}$~ergs, we have $\xi \sim 1$.  The dependence of $\xi$ on $\eta_\gamma$, $n$, and
$E_{\gamma,iso}$ is very weak.

The ``jet break'' should be independent of energy, with an achromatic
break time and the same post-break decay index, $p$, for all frequencies.
Several observational studies based primarily on breaks in optical and
radio light curves \cite{Frail01, Bloom03} have reported that jet breaks of long
GRBs typically occurred a few days after the burst, with typical jet
angles of $5\degrees - 10\degrees$, and
showed that the $\gamma$-ray energies,
corrected for the jet collimation, are tightly
clustered around $10^{51}$~ergs.  This implies a regulating mechanism that
produces similar radiation output from events thought to originate
from the gravitational collapse of massive stars of widely varying
mass, and raises the possibility that GRB afterglows may be of
cosmological use as ``standard candles''.
However, X-ray light curves from \swift\ do not exhibit similar
behavior.  We review the observational situation regarding the X-ray
light curves and discuss the implications.

\section{X-ray Jet Breaks observed by the XRT}

As of 7 October 2006, the X-ray Telescope (XRT)
\cite{Burrows05} on the {\it Swift} satellite \cite{Gehrels04} had
observed 145 ``long'' GRBs and 16 ``short'' GRBs.  
The XRT detected 143 of the ``long'' bursts and
followed their afterglows until they faded into the background,
typically at a count rate of $\sim 5 \times 10^{-4}$~s$^{-1}$.
We consider the sample of 147 of these X-ray light curves collected by the 
{\it Swift} XRT between 1 April 2005 and 7 October 2006.  Prior to 1
April 2005 {\it Swift} was primarily collecting calibration data, and
afterglow observations extending past a few days were generally not
possible.  Long-term monitoring of X-ray afterglows became common once
normal operations started.  
This data set provides the most detailed, uniform sample of GRB afterglows
available, with observations usually lasting for at least several days and often
for several weeks after the burst, far longer than the
slope changes attributed to jet breaks in earlier works.
One surprise has been the paucity of jet breaks in these X-ray light
curves.  
Based on previous work \cite{Frail01,Bloom03}, we had
expected typical X-ray light curves to break at $\sim T_0+1$~day
to $\sim T_0+4$~days to a steeper fall-off of $\alpha\sim 2.0-2.5$.
This behavior is rarely seen in the X-ray light curves from \swift.
While steepening breaks {\it are} seen in the X-ray light curves, they
usually occur at $\sim 0.1$~days post-burst and break to slopes typical
of pre-jet break (spherical) afterglows, and at least some of
these are not accompanied by similar breaks in the optical
\cite{Panaitescu06}.  It seems unlikely that these early breaks can be
attributed to jet breaks, and they have instead been attributed to the
end of an energy injection phase that often causes a flat ``plateau'' in the
X-ray light curves at typical time scales of several hundred seconds
\cite{Nousek06}.

\begin{figure}
     \centering
     \parbox{2.55in}{
    \includegraphics[width=2.4in,angle=90,bb=20 0 550 680,clip]{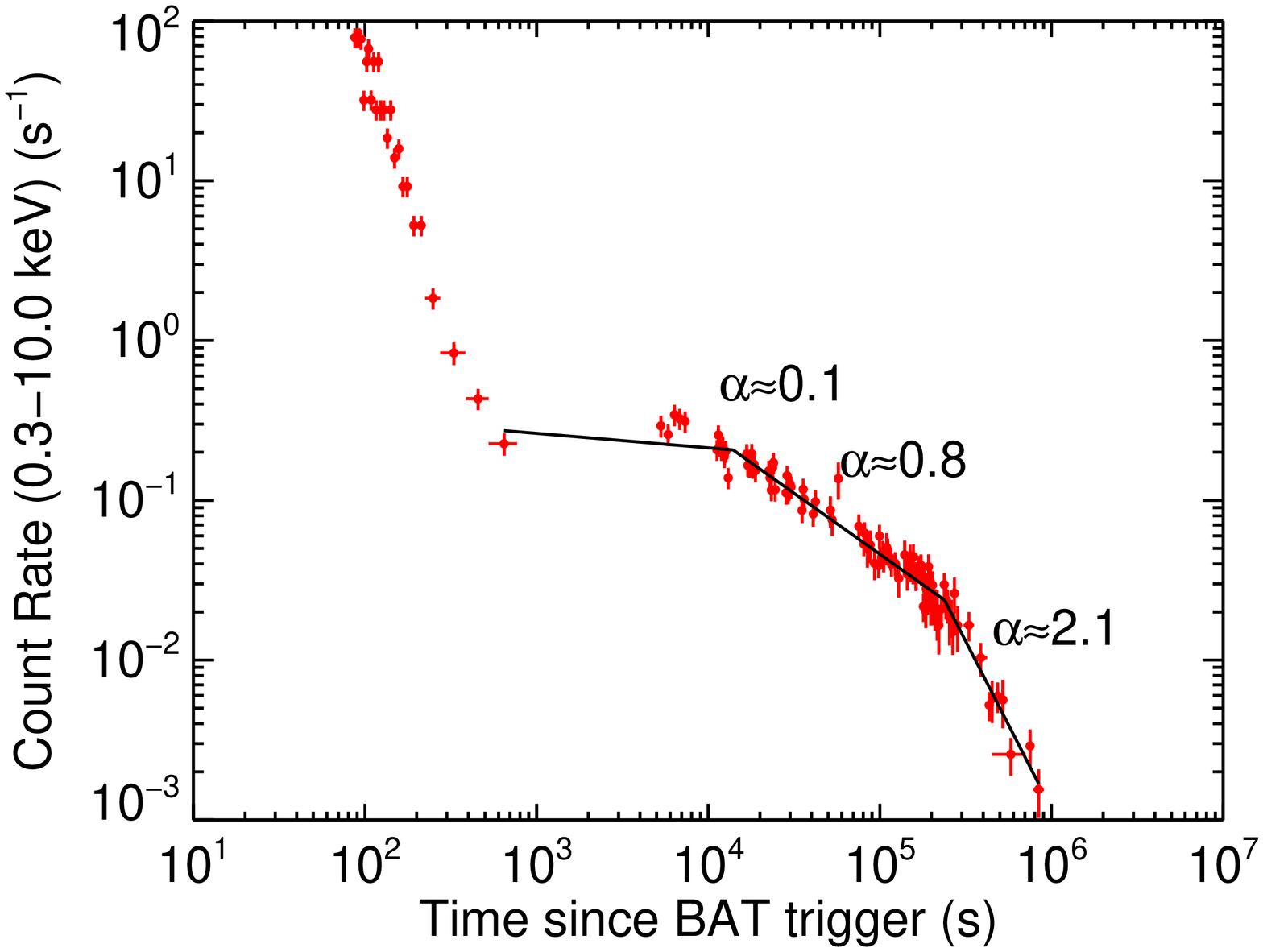}}
\hfill
     \parbox{2.55in}{
     \includegraphics[width=2.4in,angle=90,bb=20 0 550
     680,clip]{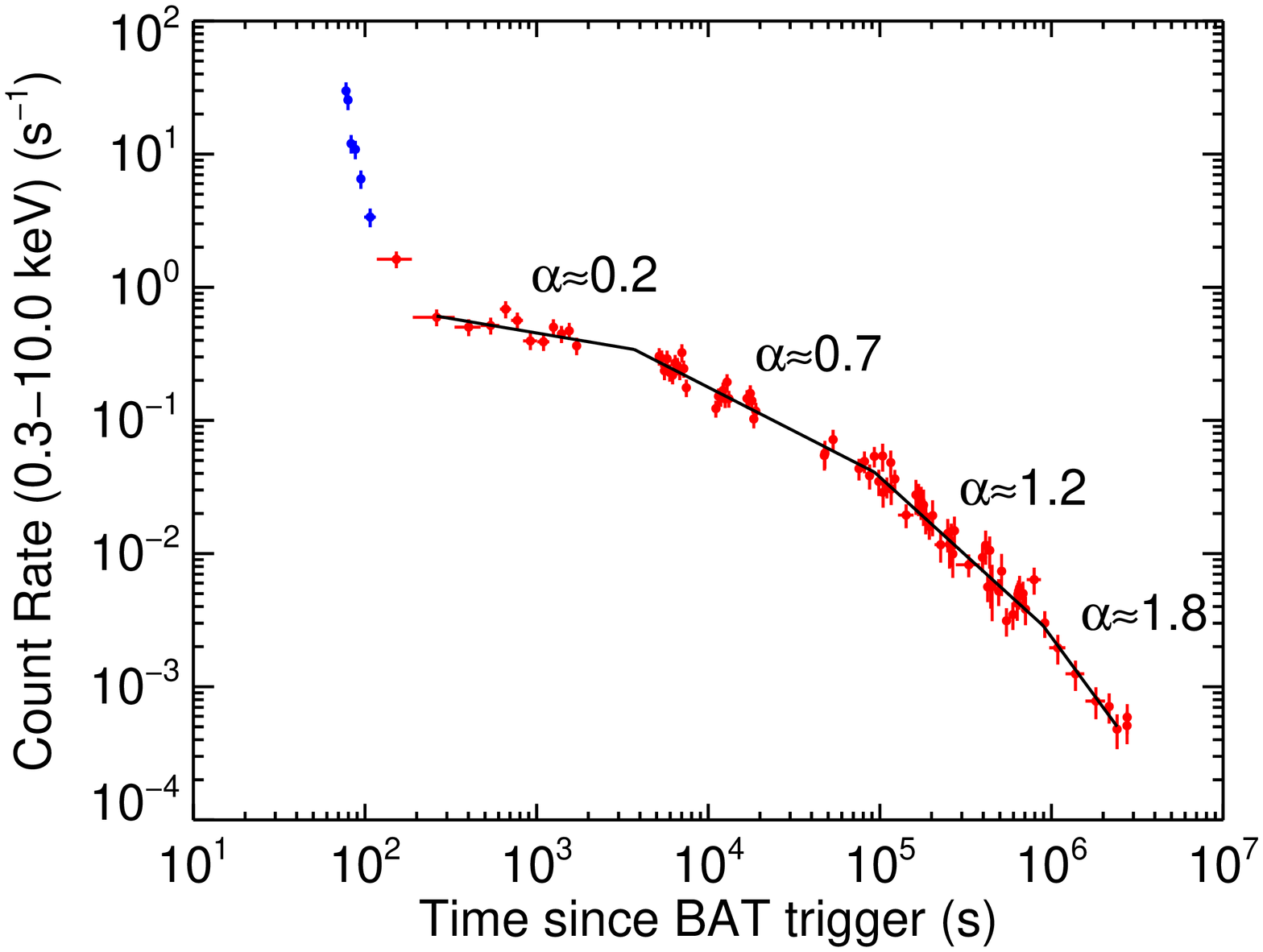}}
   \vspace{-0.2in}
   \caption{{\it Swift}/XRT light curves of GRB\,050315 (left;
     \cite{Vaughan06}) and GRB\,060428A (right). The final
     breaks in the light curves (at $\sim 240$~ks, to a slope of $\sim 2.1$
     for 050315; and at $\sim 890$~ks, to a slope of $\sim 1.8$ for
     060428A) may represent jet breaks.}
   \label{fig:050315}
   \vspace{-0.25in}
   \end{figure}

Two examples of XRT light curves with possible jet breaks 
are shown in Figure~\ref{fig:050315}.
Both of these bursts show all of the X-ray light curve segments
that are now recognized as ``canonical'' \cite{Nousek06}, although
very few GRBs actually present all of these phases.
The light curves can be represented as a series of broken power laws
with superimposed flares, with each segment described by a decay slope
$\alpha_x$ given by $F_x \propto t^{-\alpha_x}$.
The slope during the spherical expansion part of the light curve is
typically $\alpha_x \sim 1.3$ with considerable scatter.  The slope
following the jet break is expected to be $\alpha_x = p \sim 2.5$,
again with considerable scatter.  The bursts shown in
Fig.~\ref{fig:050315} have final slopes
$\alpha_x \sim 2.1$ and $1.8$, consistent with jet breaks.  
While the break time for GRB\,050315, about
2.8~days, is typical of previously reported jet break times, 
the break found in GRB\,060428A, at $\sim 10.3$~days after the
burst, is somewhat later than most previously reported jet breaks.

\begin{figure}
     \centering
     \parbox{2.55in}{
    \includegraphics[width=2.4in,angle=90,bb=20 0 550 680,clip]{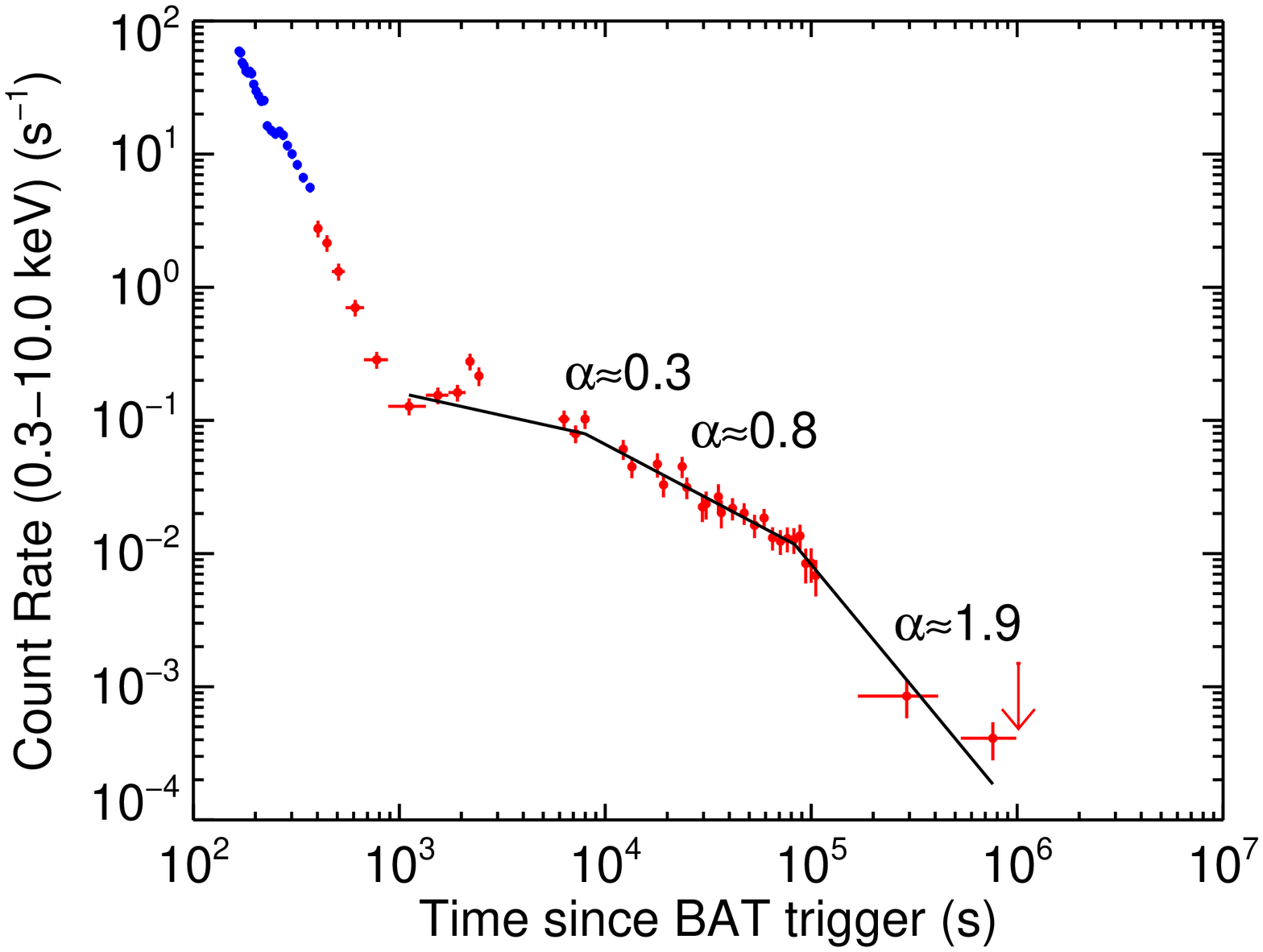}}
\hfill
     \parbox{2.55in}{
    \includegraphics[width=2.4in,angle=90,bb=20 0 550 680,clip]{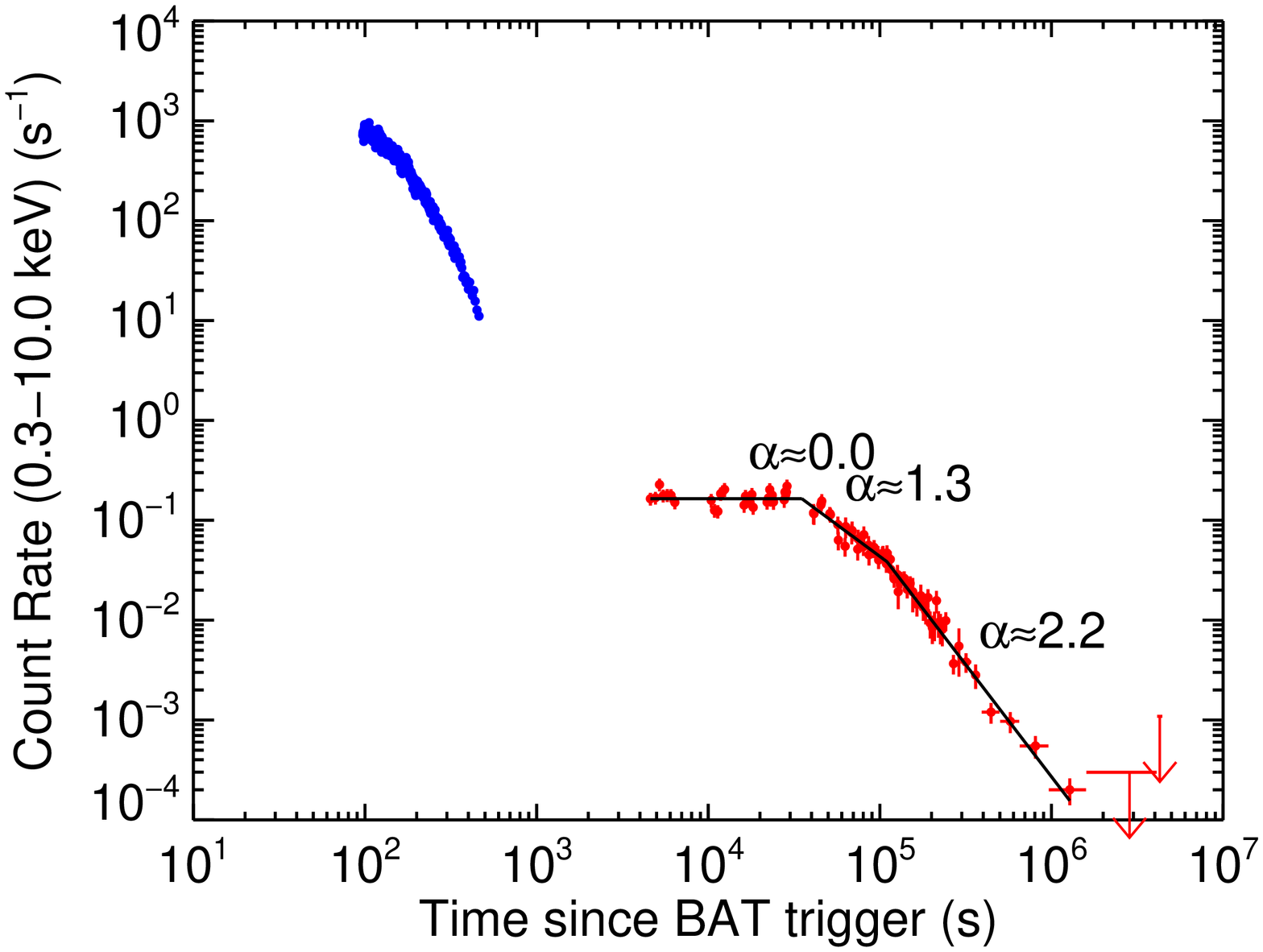}}
   \vspace{-0.2in}
   \caption{{\it Swift}/XRT light curves of GRBs\,050814 (left) and
     060614 (right; \cite{Mangano07}).  The final
     breaks in the decay slopes (at $\sim 85$~ks, to a slope of 1.9
     for 050814; and at $\sim 110$~ks, to a slope of 2.2 for 060614) are typical
     of jet breaks.  GRB\,060614 is one of the best examples of a jet
     break, with an achromatic break in the optical and X-ray bands \cite{Mangano07}.
     }
   \label{fig:jet_breaks}
   \vspace{-0.2in}
   \end{figure}

Two more potential jet breaks are shown in Fig.~\ref{fig:jet_breaks}.
GRB\,050814 breaks from a shallow plateau phase with $\alpha \approx
0.8$ to a post-break slope of 1.9.  Like GRB\,050315, the break occurs
in the expected time-frame based on previous work: about 1.0~days in
the observer frame in
this case (though with a redshift of $z=5.3$ \cite{Jakobsson06} this
break occurs at only 13~ks in the GRB rest frame).
For $E_{\gamma,iso} \sim 1.8 \times 10^{53}$ (Sakamoto, private
communication), we obtain a jet angle of $\theta_j \sim
3\degrees~n^{1/8}$ from Eq. 1 (for $\eta_\gamma=0.2$).
GRB\,060614 is one of the best examples of a Swift burst with a jet
break, having an achromatic break at about 1.3 days in both X-ray and
optical bands, and breaking to a slope of $\sim 2.2$ \cite{Mangano07}.  We estimate a
jet angle for this burst of about $10 \degrees~n^{1/8}$, and $E_\gamma
\sim 3.6 \times 10^{49}$~erg.  It is not clear whether this
burst is a subluminous ``long'' GRB with no SN component 
\cite{Gehrels06,GalYam06,Fynbo06a,DellaValle06} or a highly energetic ``short''
GRB \cite{Zhang07}.

\section{Long-term X-ray afterglows without jet breaks}

X-ray light curves with jet breaks, however, are decidedly in the minority.  In a detailed
study of 107 \ X-ray afterglow light curves \cite{Willingale07}, only a few percent are found
that satisfy the closure relations between the decay slope and the
spectral slope expected for post-jet break light curves (e.g. \cite{Zhang04}).  
The vast majority have no evidence for a break in the light curve with
the characteristics expected for jet breaks.  To illustrate the
problem, we present examples of X-ray light curves of GRBs with long
observation times.  Although this is a small sample of a much larger
collection of X-ray light curves, these are typical of the overall
group.  We make the assumption that the jet break occurs after the last data
point in these cases.

\begin{figure}
     \centering
     \parbox{2.55in}{
    \includegraphics[width=2.3in,angle=90,bb=20 0 550 680,clip]{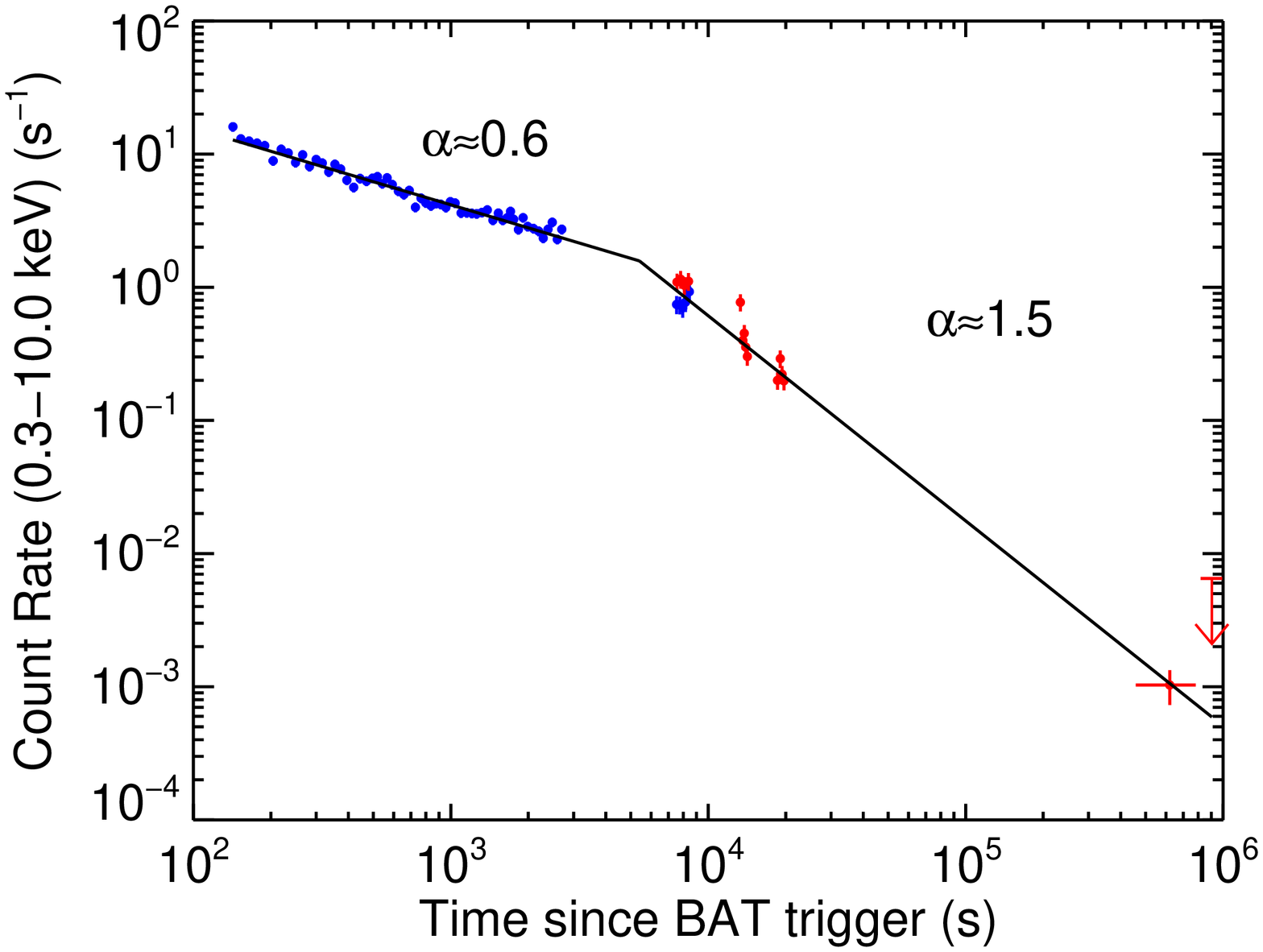}
   \caption{{\it Swift}/XRT light curve of GRB\,050401.  The final
     break in the X-ray light curve is at about 4.9 ks, to a final decay
     slope of 1.5 \cite{dePasquale06}, typical for a normal
     ``spherical'' (pre-jet break) afterglow and much
     flatter than expected for a post-jet break light curve.}
   \label{fig:050401}}
   \hfill
     \parbox{2.55in}{
    \includegraphics[width=2.3in,angle=90,bb=20 0 550 680,clip]{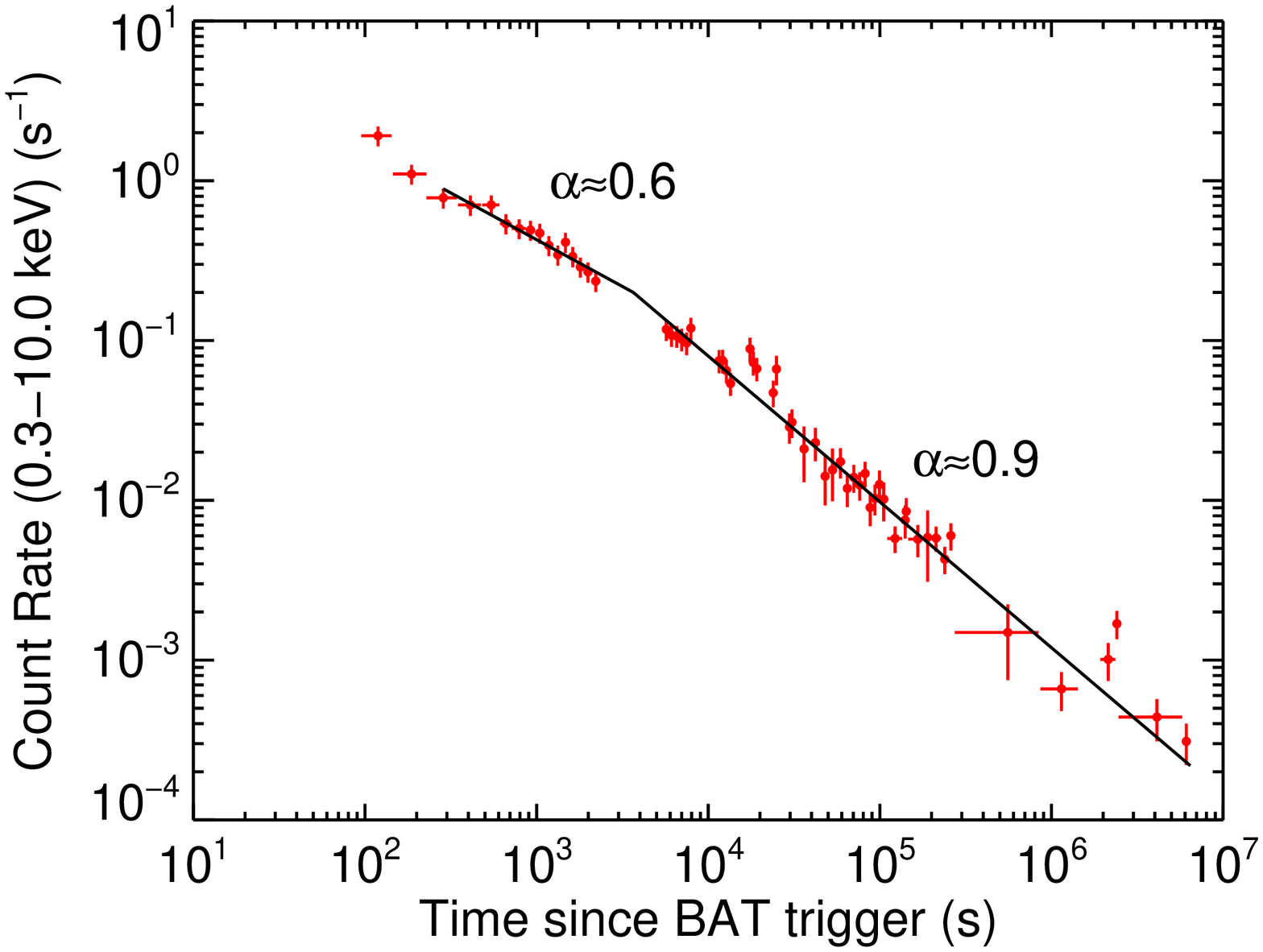}
   \caption{{\it Swift}/XRT light curve of GRB\,050416A 
     \cite{Mangano06}.  The final
     break in the light curve is at about 1.5 ks, to a final decay
     slope of 0.9, a bit shallow for a normal afterglow, and far
     flatter than expected for a post-jet break light curve.}    
   \vspace{0.2in}
   \label{fig:050416A}}
   \vspace{-0.2in}

   \end{figure}

We begin with the X-ray light curve of GRB\,050401 (Fig.~\ref{fig:050401}).
Although this light curve has several large gaps, it extends to about
800~ks with no evidence for any break after about 7~ks.  The final 
break in the light curve is at 4.9~ks, and the final slope is $1.46
\pm 0.07$ \cite{dePasquale06}.  This is a typical slope for a
spherically-expanding afterglow, but is far too shallow to satisfy the
closure relations for
the post-jet break case \cite{dePasquale06}.  We conclude that any jet
break must occur after $\sim 800$~ks.  With a
redshift of 2.9 and $E_{iso}=3.5 \times 10^{53}$~ergs \cite{dePasquale06}, we find a lower limit
on the opening angle of $\theta_j > 7\degrees~n^{1/8}$.  Because of the relatively
large redshift and high isotropic energy, this rather late limit to a
jet break time is compatible with typical 
opening angles from \cite{Frail01,Bloom03}.  

GRB\,050416A (Fig.~\ref{fig:050416A}) puts very strong constraints on a
jet break time.  This burst has one of the longest afterglow followups
to date with the \swift/XRT, with observations continuing for 74 days
after the burst \cite{Mangano06}.  With more sensitive reanalysis of
this light curve, we detect the afterglow up to 70 days
post-burst, with a slope of $\alpha=0.9$ from 1.45~ks to 6~Ms (Fig.~\ref{fig:050416A}).  This
is somewhat flatter than expected for a spherical afterglow
($\alpha\sim 1.1$ for this case).  Using Eq.~(1), we find that
$\theta_j > 42\degrees~n^{1/8}$.

\begin{figure}
     \centering
     \parbox{2.55in}{
    \includegraphics[width=2.3in,angle=90,bb=20 0 550 680,clip]{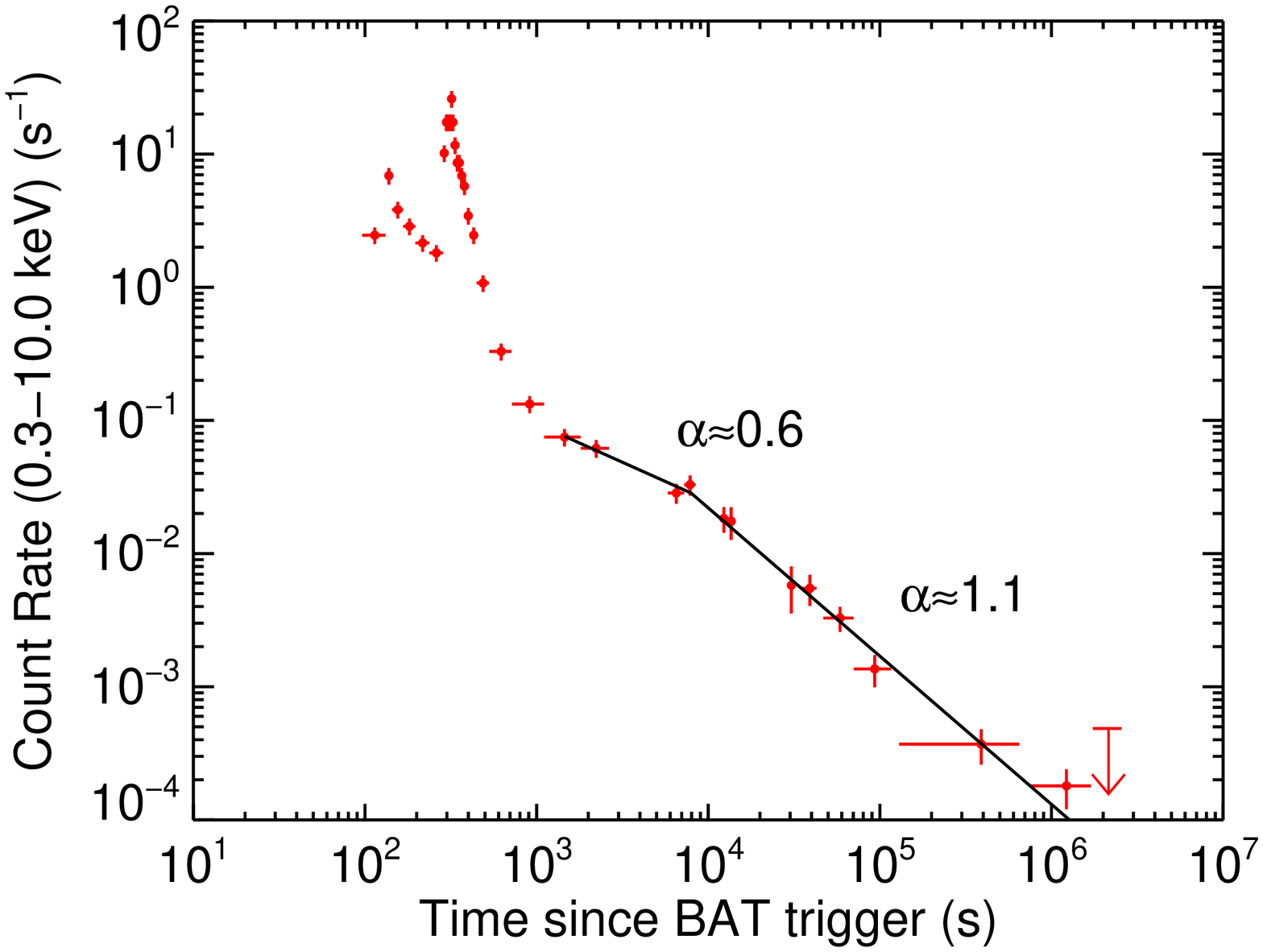}
   \caption{{\it Swift}/XRT light curve of GRB\,050607 \cite{Pagani06}. The final
     break in the light curve is at about 8 ks, to a final decay
     slope of 1.1.
   }
   \label{fig:050607}}
   \hfill
%
     \parbox{2.55in}{
    \includegraphics[width=2.3in,angle=90,bb=20 0 550 680,clip]{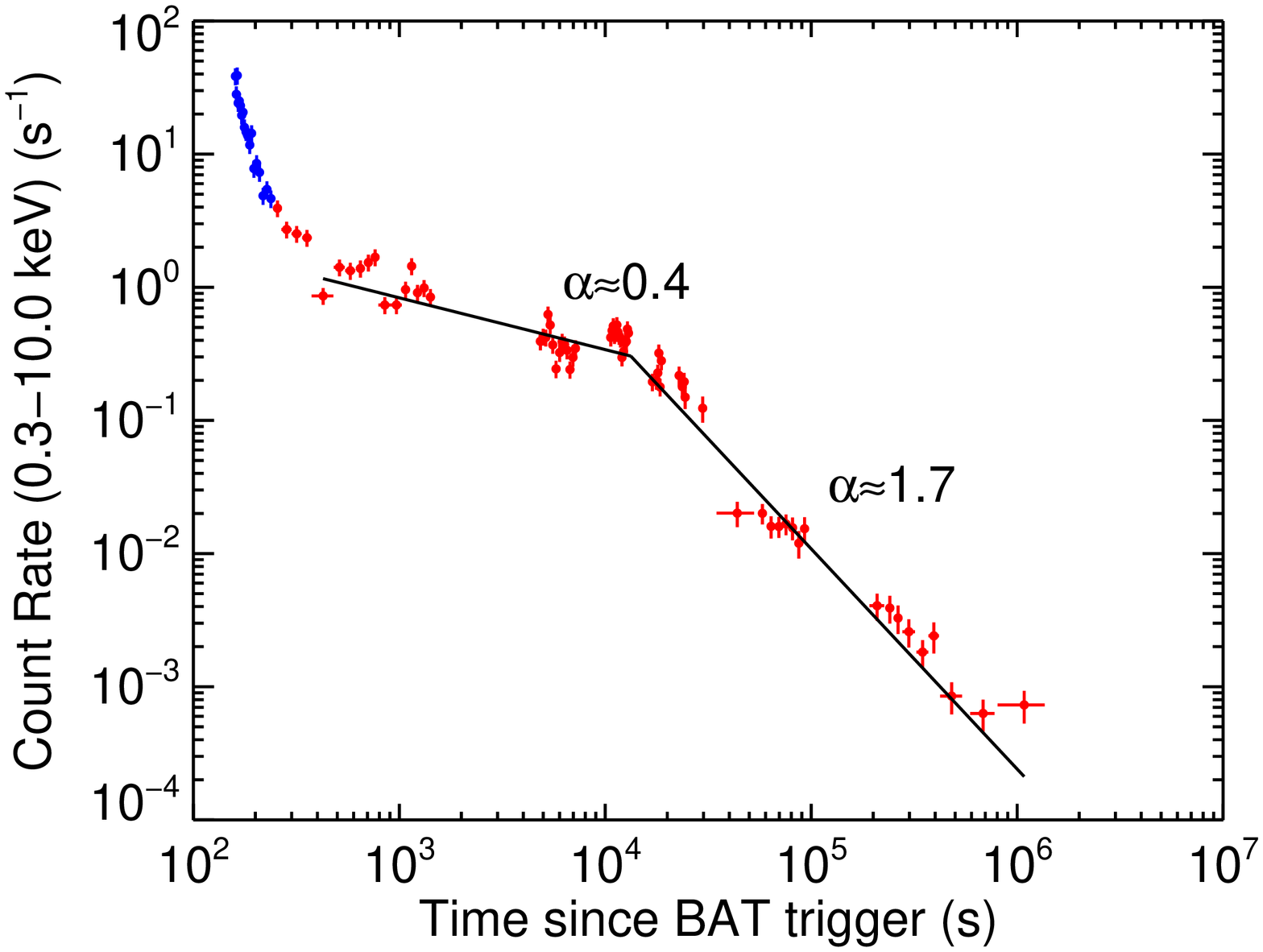}
   \caption{{\it Swift}/XRT light curve of GRB\,050803. The final
     break in the light curve is at $\sim 13$~ks, to a final decay
     slope of 1.7.}
   \label{fig:050803}}

   \end{figure}

GRB~050607 (Fig.~\ref{fig:050607}) has a typical XRT light curve, with strong flares at early
times, a plateau phase, and then a normal afterglow phase extending
until 1.7~Ms post-burst, when it becomes undetectable.  This last
phase has a slope of $\alpha=1.1$ and fits the closure relation
expected for a standard afterglow propagating into a uniform density
ISM, with electron index $p\sim2.2$ \cite{Pagani06}.
The redshift is not known for this burst, so the jet angle cannot be
determined directly.
We use Eq.~2 to obtain an estimate of $\theta_j >
11\degrees \xi$.

GRB\,050803 (Fig.~\ref{fig:050803}) has a somewhat erratic light curve.  After $\sim 15$~ks
the light curve breaks to a slope of $\alpha=1.7$, with considerable
scatter that may represent small flares superimposed on the afterglow.
This slope continues for about 13~days post-burst, giving us an
estimate for the jet opening angle of
$\theta_j > 10\degrees \xi$.

   \begin{figure}
     \centering
     \parbox{2.55in}{
    \includegraphics[width=2.4in,angle=90,bb=20 0 550 680,clip]{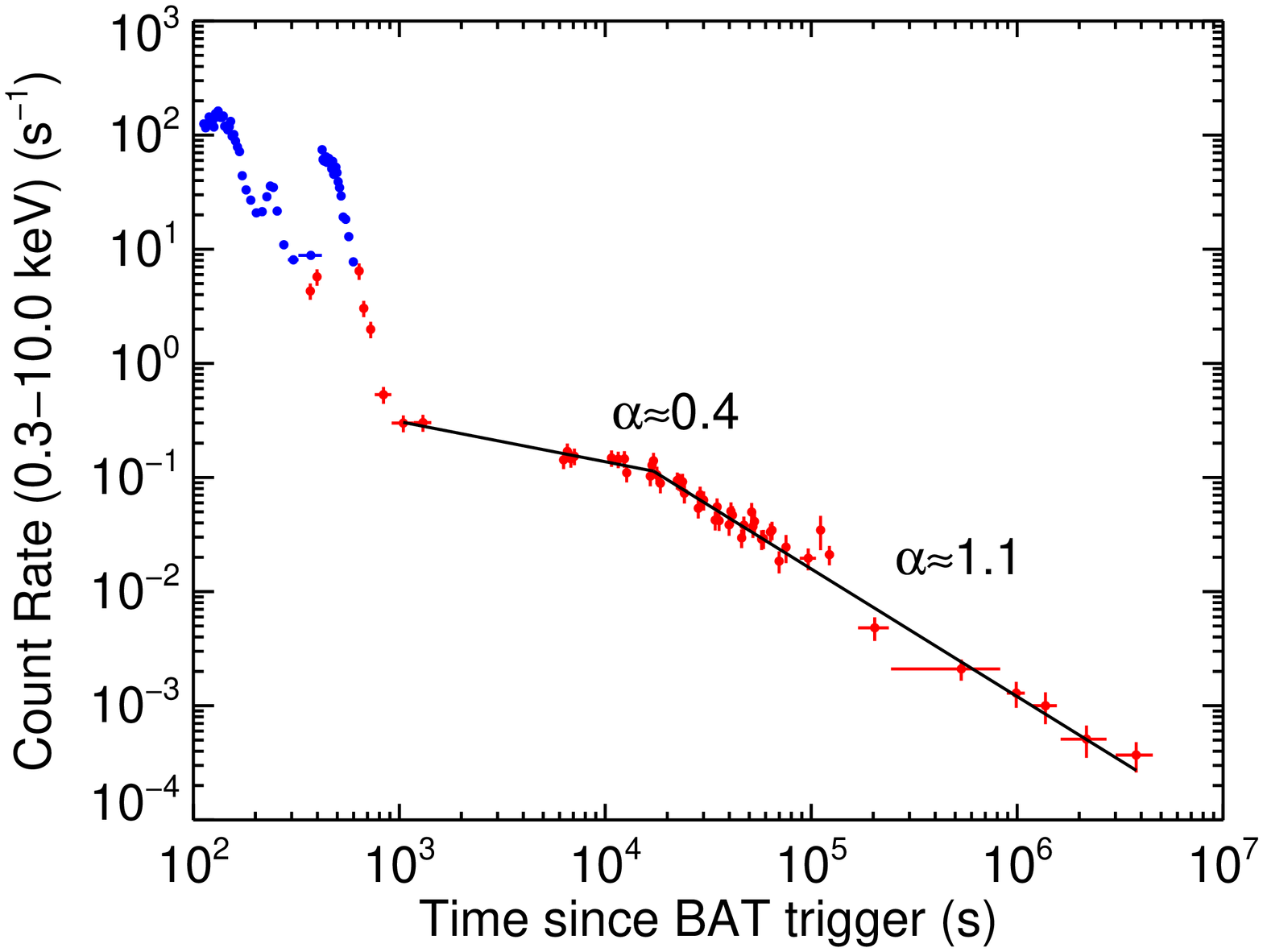}}
   \hfill
     \parbox{2.55in}{
    \includegraphics[width=2.4in,angle=90,bb=20 0 550 680,clip]{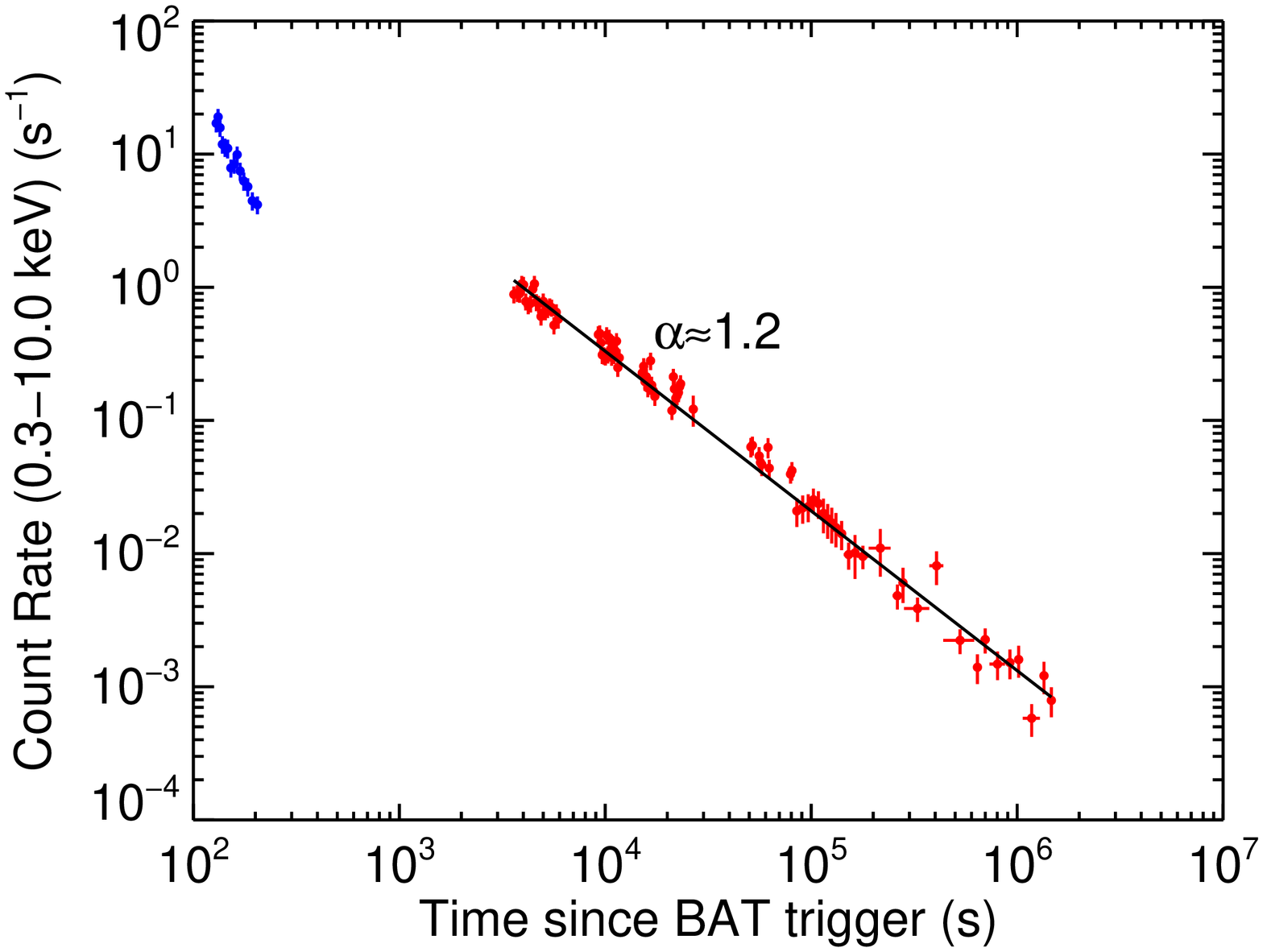}}
   \vspace{-0.2in}
   \caption{{\it Swift}/XRT light curves of GRBs\,050822
     (left; \cite{Godet06}) and 051109A (right). Both light curves decay at slopes typical
     for spherical afterglows, and neither light curve has evidence
     for a jet break to times later than 16~days.}
   \label{fig:051109A}
   \vspace{-0.2in}
   \end{figure}
GRB\,050822 (Fig.~\ref{fig:051109A}; \cite{Godet06}) has flares at early
times, a very flat plateau phase, and a standard decay after $\sim
15$~ks with a slope $\alpha=1.1$.  
There is no evidence for a jet break before 52~days post-burst.
This burst has no optical counterpart and hence no redshift
measurement, but \cite{Godet06} show that such a late break is
unexpected for any redshift less than 10.
We find a jet angle of $\theta_j > 16\degrees \xi$ for this afterglow.

One can infer a plateau phase for GRB\,051109A (Fig.~\ref{fig:051109A})
during the first orbital gap, followed by a simple power law decay
with $\alpha=1.2$ until 16 days post-burst, with low-level flaring superimposed on the
power law decay.
This burst has a redshift of 2.346 \cite{Quimby05} and $E_{iso} = 5
\times 10^{52}$\,ergs \cite{Golenetskii05}, from which we can determine a minimum opening angle of
$\theta_j > 12 \degrees~n^{1/8}$.

   \begin{figure}
     \centering
     \parbox{2.55in}{
    \includegraphics[width=2.4in,angle=90,bb=20 0 550 680,clip]{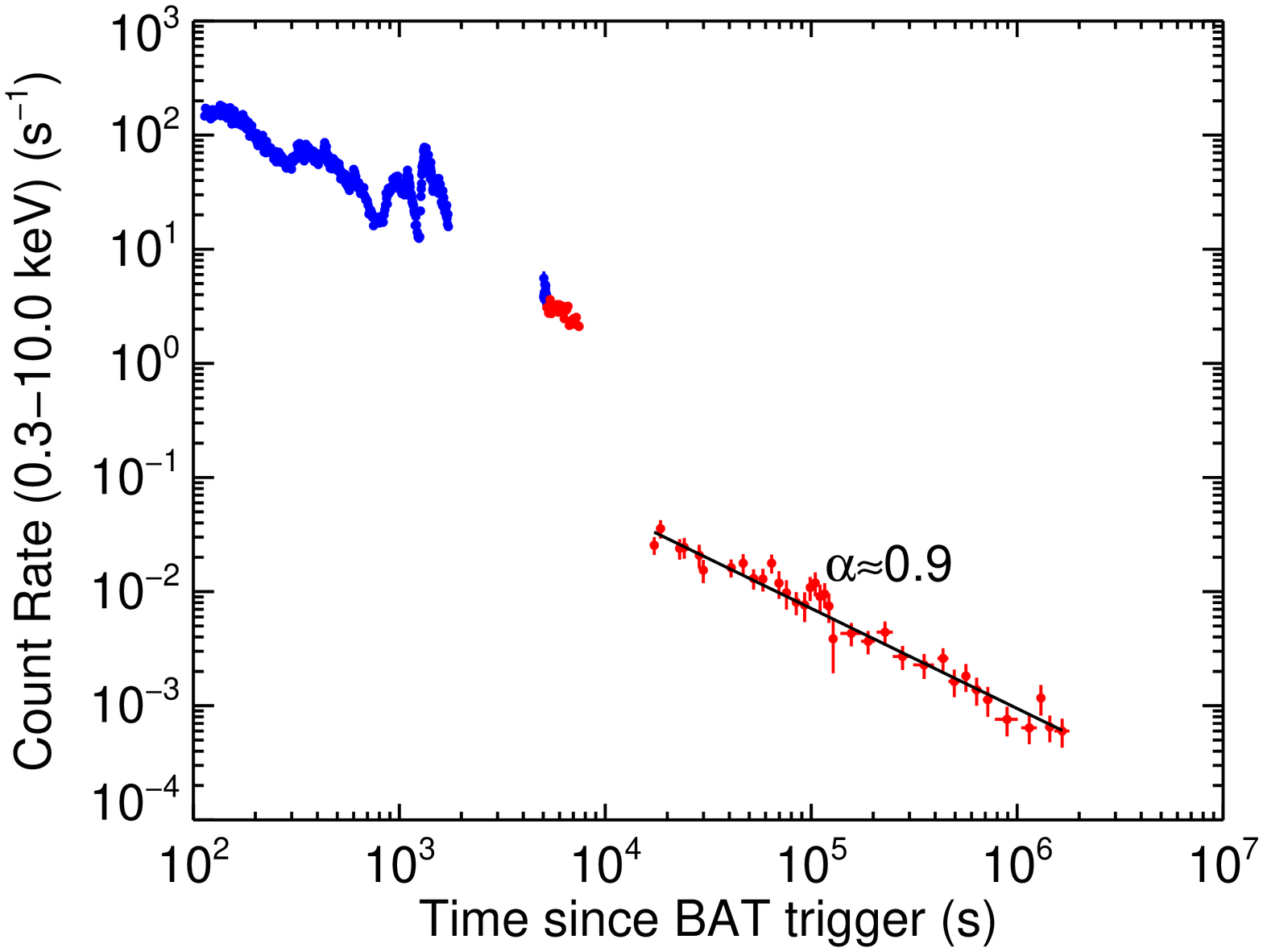}}
   \hfill
     \parbox{2.55in}{
    \includegraphics[width=2.4in,angle=90,bb=20 0 550 680,clip]{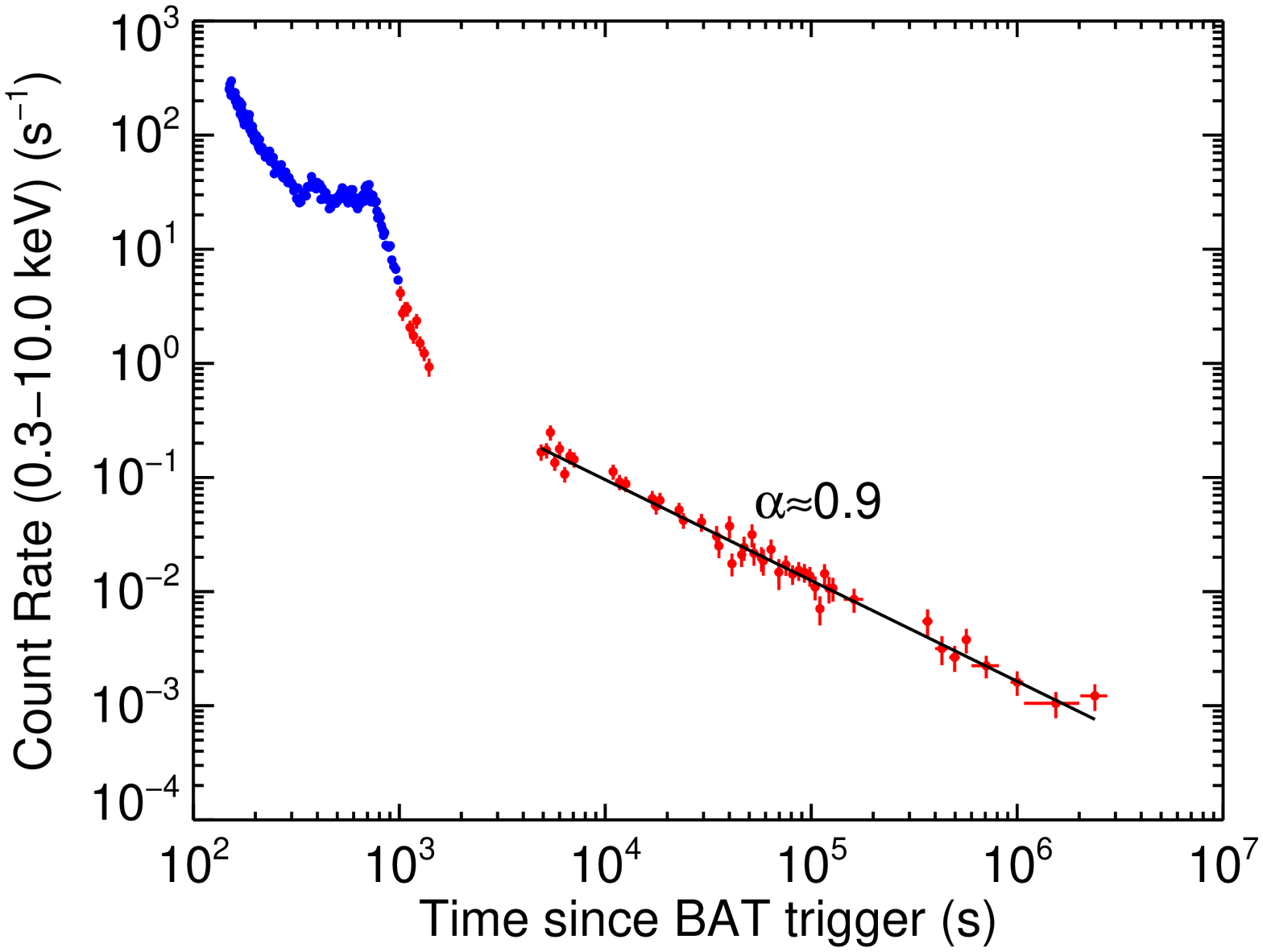}}
   \vspace{-0.2in}
   \caption{{\it Swift}/XRT light curves of GRB\,051117A (left; \cite{Goad06})
     and GRB\,060202 (right). Both light curves display considerable flaring at
     early times, but then settle down to a simple power law afterglow
     decay with slopes of 0.9, lasting from $\sim 10^4$~s until $>2 \times 10^6$~s after the
     burst trigger.  } 
   \label{fig:060202}
   \vspace{-0.2in}
   \end{figure}
GRB\,051117A (Fig.~\ref{fig:060202}; \cite{Goad06}) and GRB\,060202
both have long light curves, beginning with substantial flaring and
then settling down to simple power law decays at late times.  Both
GRBs have a rather flat final decay slope of 0.9 from $\sim 10^4$~s until 
the afterglow becomes undetectable by the XRT.  The afterglow
continues until $\sim 20$ days post-burst for 051117A, and until $\sim 29$
days post-burst for 060202.
Neither burst has a redshift measurement, so we use Eq.~2 to estimate their jet
angles as $\theta_j > 11\degrees \xi$ for 051117A and $\theta_j >
13\degrees \xi$ for 060202.

   \begin{figure}
     \centering
     \parbox{2.55in}{
     \includegraphics[width=2.7in,bb=57 8 872 582,clip]{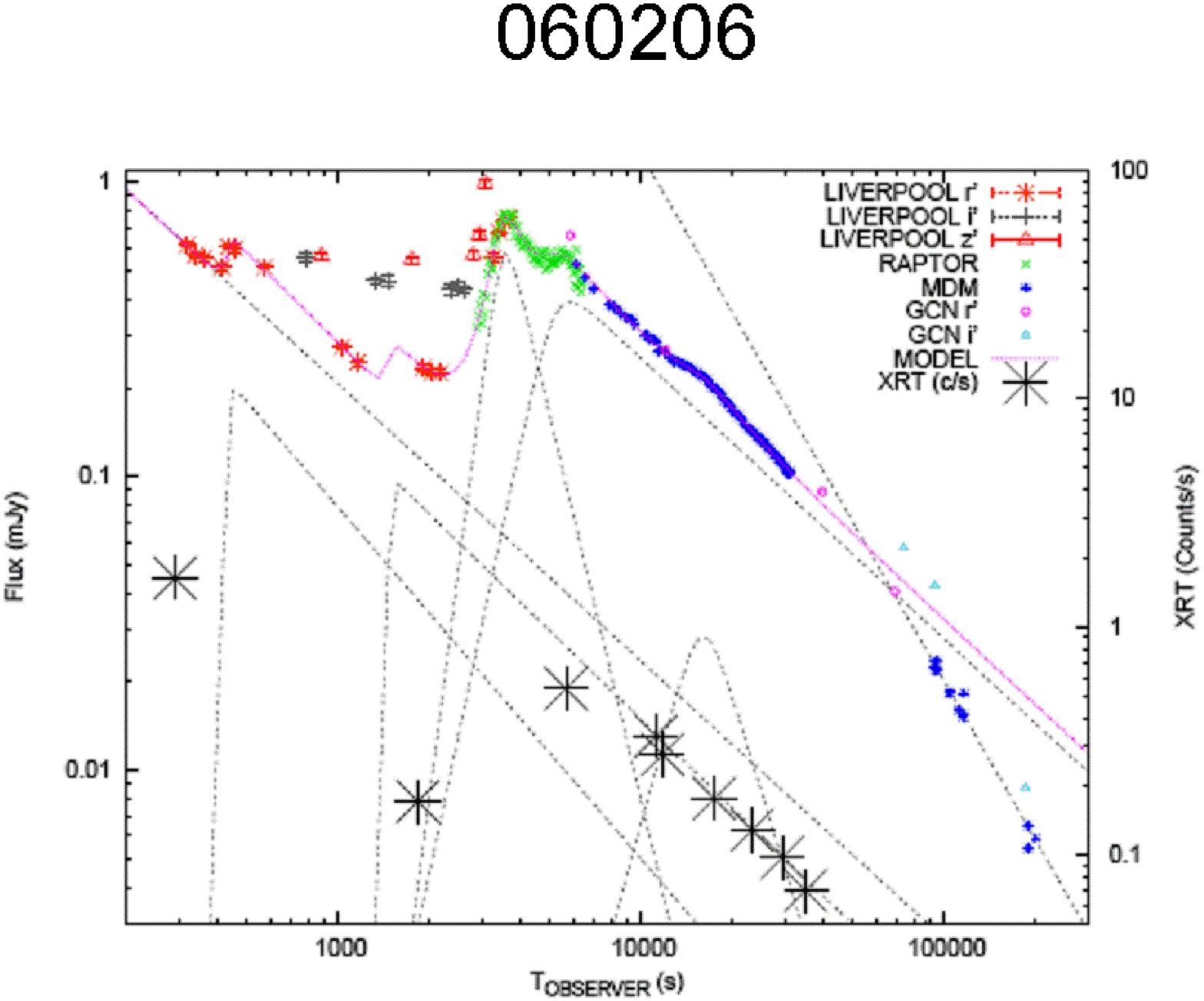}}
   \hfill
     \parbox{2.55in}{
    \includegraphics[width=2.4in,angle=90,bb=20 0 550 680,clip]{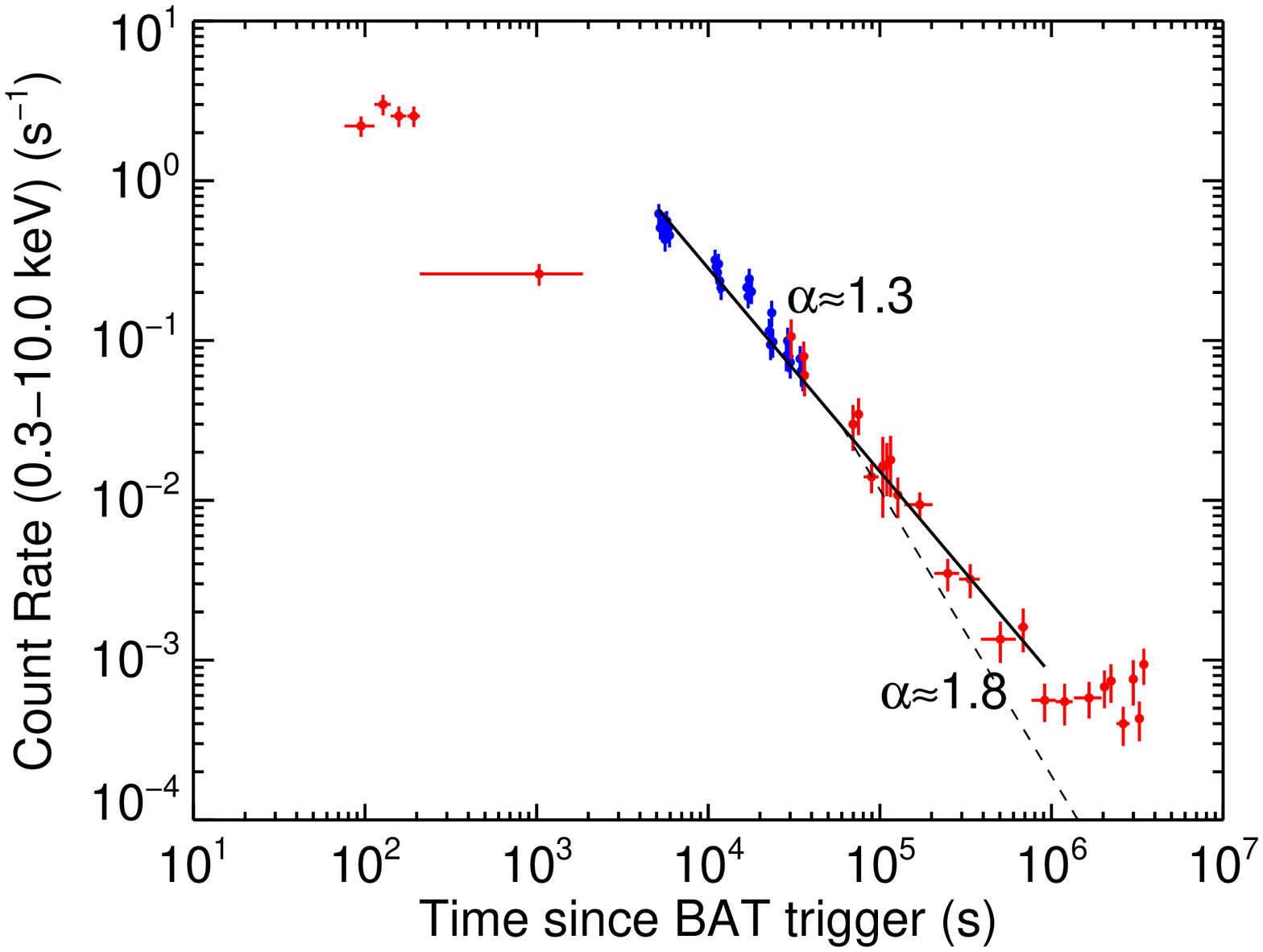}}
   \vspace{-0.2in}
   \caption{Left: Optical light curve of GRB\,060206 (from
     \cite{Monfardini06}).  There is a clear break in the optical light
     curve at $\sim 60$~ks.  Right: {\it Swift}/XRT light curve of
     GRB\,060206. Although the data are somewhat ``noisy'', perhaps
     due to small-scale flaring, there is no evidence of any break in
     the X-ray light curve after $\sim 5$~ks (until it flattens out at
     about $10^6$~s).  The dashed line
     indicates the X-ray light curve expected if the optical break at
     $\sim 60$~ks were achromatic, and is clearly a poor fit to the
     late data.}
   \label{fig:060206}
   \vspace{-0.25in}
   \end{figure}
We next consider GRB\,060206 (Fig.~\ref{fig:060206}), a high redshift burst with z=4.048
\cite{Fynbo06b} and $E_{\gamma,iso}=5.8 \times 10^{52}$~ergs \cite{Palmer06}.  
This burst had a bright, particularly
well-sampled optical light curve for the first two days of the
afterglow \cite{Monfardini06,Stanek06}.
The optical light curve breaks from a slope of $\alpha_o = 0.95 \pm
0.02$ to a steeper slope of $1.79 \pm 0.11$ at a time of $\sim
30-90$~ks \cite{Monfardini06}.  This break is interpreted as a jet break
by \cite{Stanek06}, who overplot the X-ray light curve from 60~s to
230~ks on the optical
light curve, normalized so that the highest points on the X-ray light
curve match the optical light curve in the interval 5-40~ks.  
Over this limited range, with the large scatter in the X-ray data
points, the X-ray and optical light curves agree reasonably well, which 
\cite{Stanek06} claim as evidence that the X-ray light curve follows the optical
light curve.
This would imply that the break in the optical light curve is achromatic.  
However, examination of the entire X-ray light
curve makes it clear that there is no evidence for any change in the overall X-ray slope of
$\sim 1.3$ between 5~ks and at least 15 days post-burst
(Fig.~\ref{fig:060206}; from 15 days to 26 days the light curve is flat,
after which the source became undetectable).  Furthermore, we
note that variability in X-ray light curves of the sort seen in this
light curve is fairly common, and it
is likely to be due to mini-flares superimposed on a relatively smooth
underlying afterglow.  This variability, combined with the large
uncertainties in the X-ray data points, can account for the apparent
agreement between the optical and X-ray light curves over the small
time interval considered by \cite{Stanek06}.
We conclude that the optical
break is a chromatic break unassociated with a jet break, in agreement
with \cite{Monfardini06}.  In our
interpretation, the X-ray light curve is unbroken from 5~ks to
at least 1.3~Ms, or 15 days post-burst.  We derive a limit of
$\theta_j > 10 \degrees~n^{1/8}$ for the jet opening angle.

   \begin{figure}
     \centering
     \parbox{2.55in}{
    \includegraphics[width=2.4in,angle=90,bb=20 0 550 680,clip]{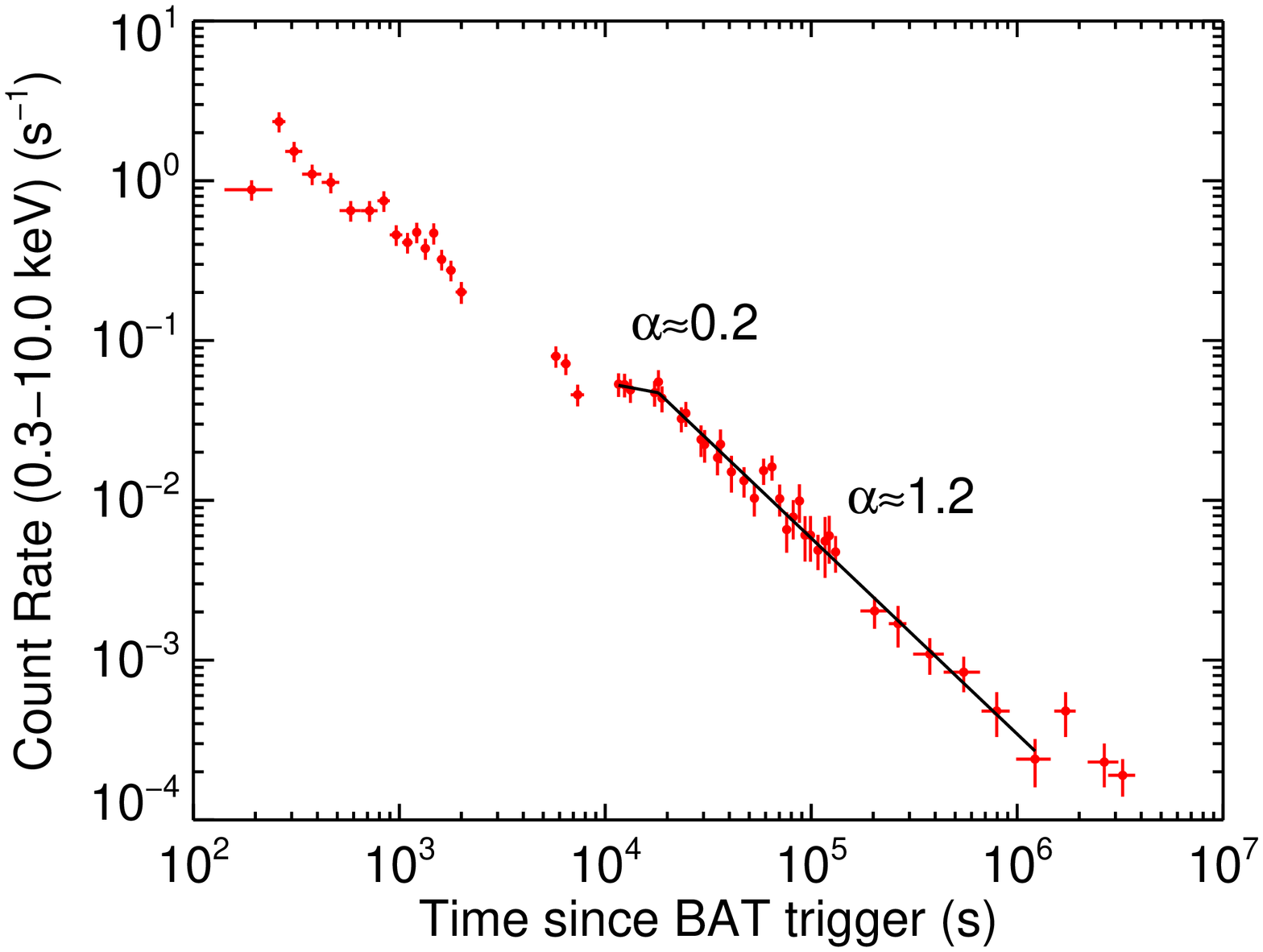}
   \caption{{\it Swift}/XRT light curve of GRB\,060319. 
     The decay slope is typical for a spherical afterglow, with no jet
     break until at least 3.6~Ms.
   }
   \label{fig:060319}}
   \hfill
%
%
     \parbox{2.55in}{
    \includegraphics[width=2.4in,angle=90,bb=20 0 550 680,clip]{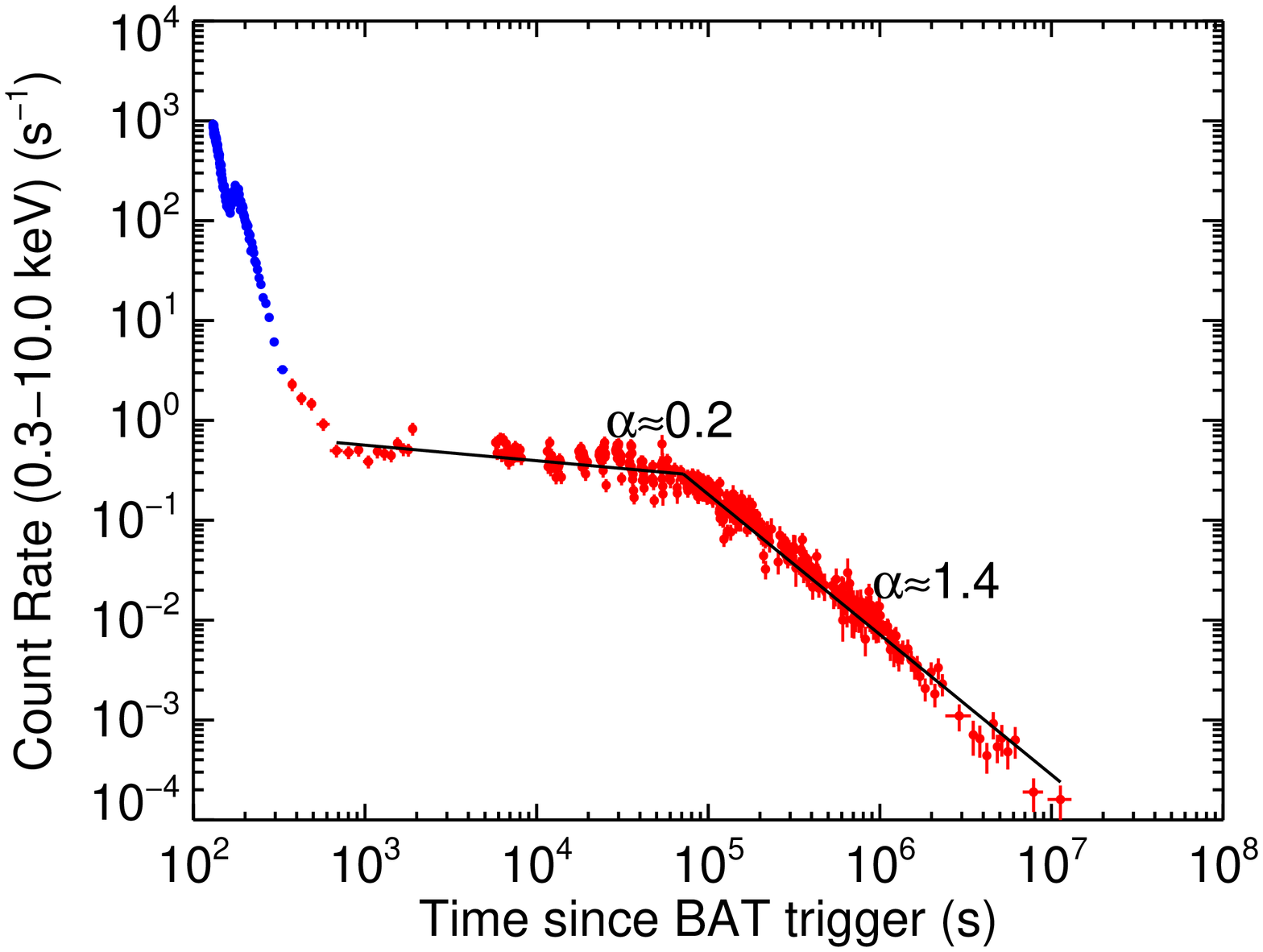}
   \caption{XRT light curve of GRB\,060729. The final
     break in the light curve is at about 60~ks, to a final decay
     slope of $\alpha= 1.4$.}
     \vspace{0.1in}
   \label{fig:060729}}
   \vspace{-0.30in}
   \end{figure}
GRB\,060319 is another exceptionally long X-ray light curve, extending
to 42 days post-burst (Fig.~\ref{fig:060319}).  The initial
decay slope is $\sim 1.2$, though with significant variability,
from 230~s to 8ks.  Between 8~ks and 18~ks there is an episode of
energy injection into the external shock, after which the
afterglow resumes a steady decay slope of $\alpha=1.2$ that
continues until 1.3~Ms, with a very late flare or energy injection
episode followed by a continuing decay until 3.6~Ms.  
This afterglow has no optical counterpart or
redshift.  We estimate the jet break opening angle to be 
$\theta_j >  15\degrees \xi$.

GRB\,060729 has the longest X-ray afterglow seen to date by \swift, decaying at a
slope of $\alpha=1.4$ for over 10~Ms \cite{Grupe06} ($\sim 81$ days in
the GRB rest frame!).
Fig.~\ref{fig:060729} shows the first 5~Ms of the light curve.  
The object was unusually bright when the observation began.
A typical steep decay with a strong flare was followed by an extremely
flat, unusually long plateau phase extending to $\sim 60$~ks.  
The final decay slope is typical for a pre-jet break spherical
outflow. With a redshift of $z=0.54$ \cite{Thoene06} and $E_{\gamma,iso}=1.6
\times 10^{52}$~ergs, we find $\theta_j > 38\degrees~n^{1/8}$ for this
afterglow.

\begin{figure}
     \centering
     \parbox{2.55in}{
    \includegraphics[width=2.4in,angle=90,bb=20 0 550 680,clip]{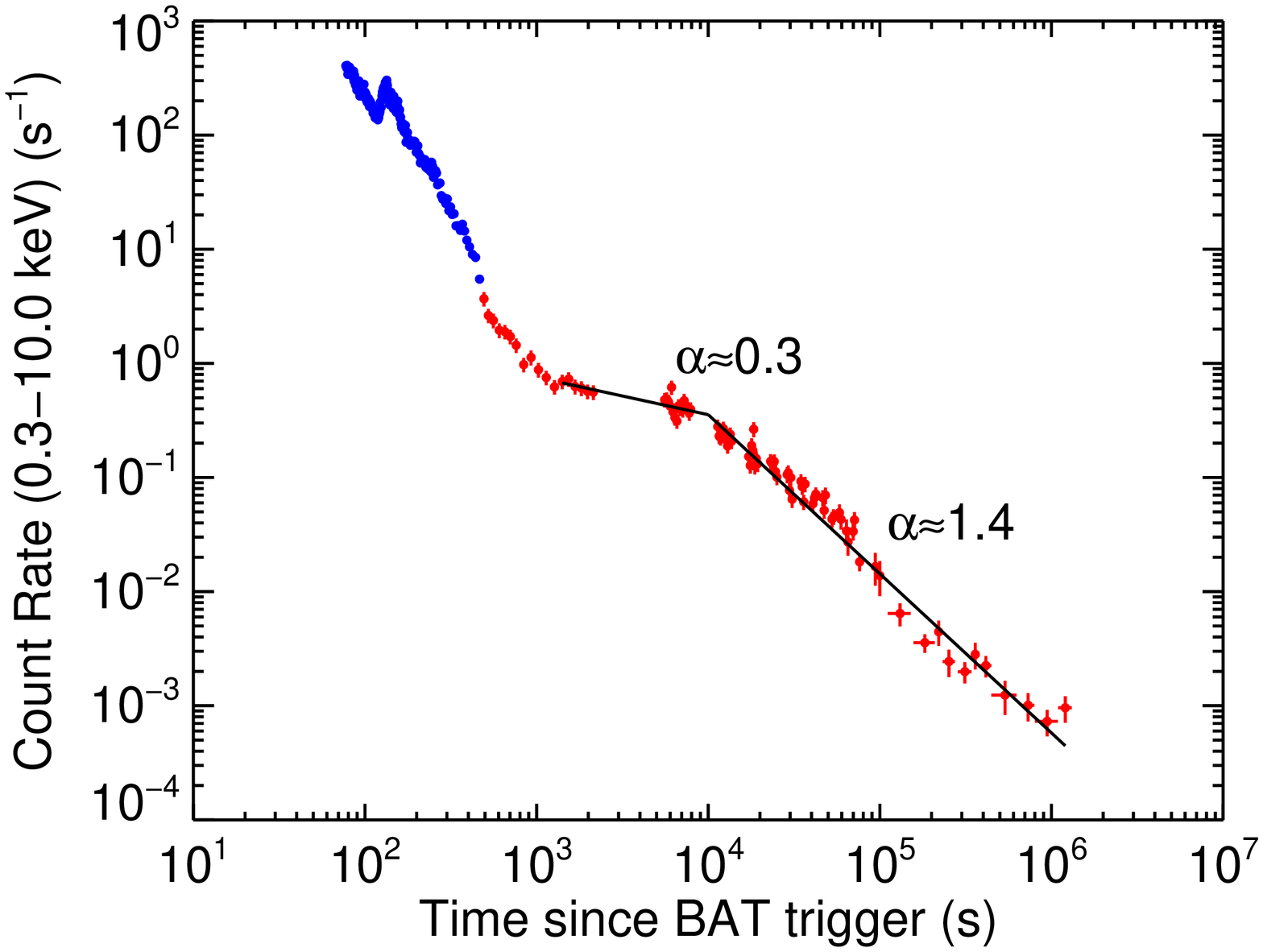}
   \caption{{\it Swift}/XRT light curve of GRB\,060814, showing the
     typical decay phases: rapid decline with flares, plateau, and
     power-law decay with slope of $\alpha \sim 1.4$, with some
     late-time flares superimposed.
   }
   \label{fig:060814}}
   \hfill
     \parbox{2.55in}{
    \includegraphics[width=2.4in,angle=90,bb=20 0 550 680,clip]{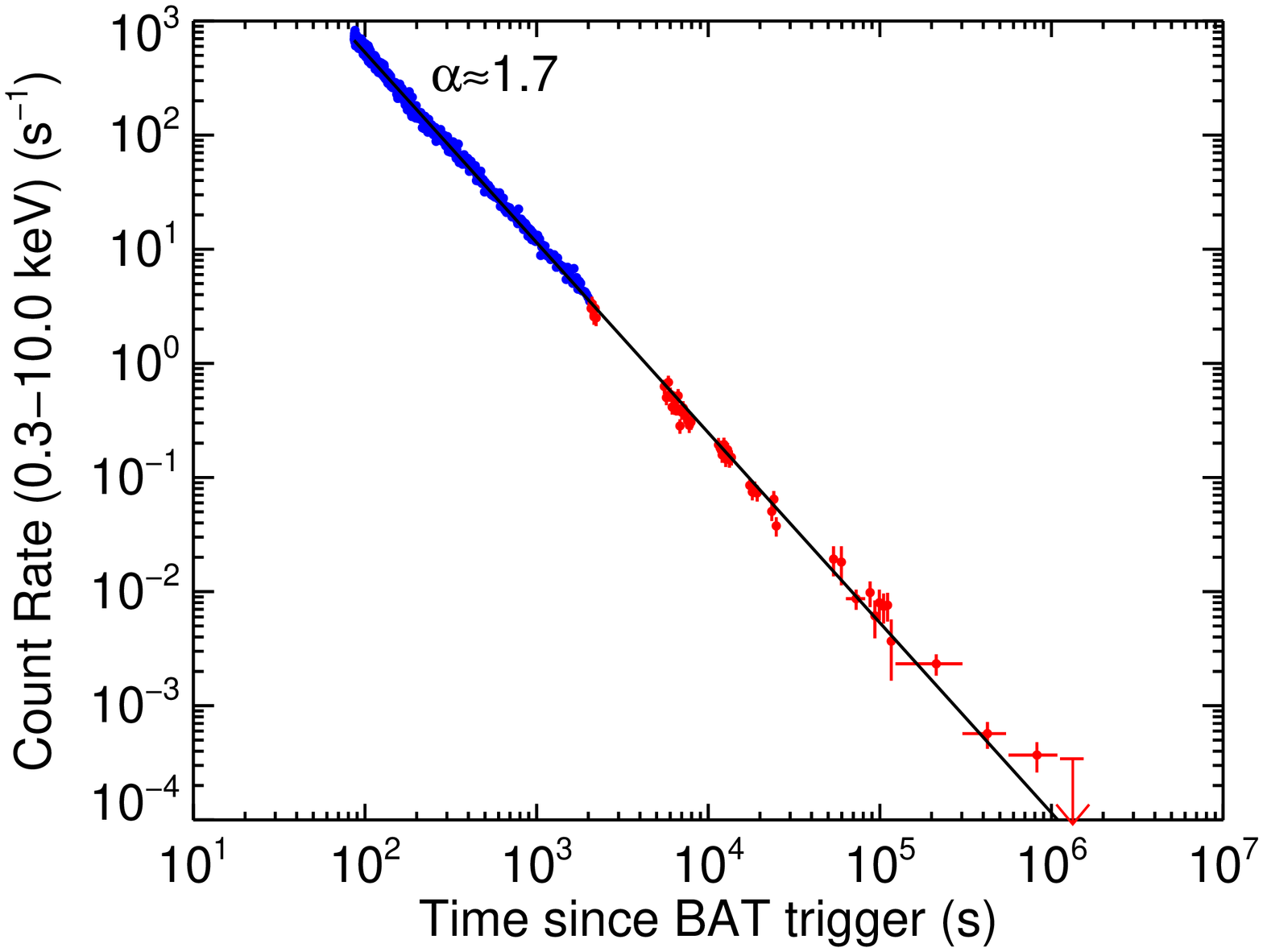}
   \caption{The unusual {\it Swift}/XRT light curve of GRB\,061007
     \cite{Schady06}, which consists of a single power-law decay
     ($\alpha=1.66$) for
     over 1~Ms.  }
     \vspace{0.2in}
   \label{fig:061007}}
   \vspace{-0.2in}
   \end{figure}

The light curve of GRB\,060814 (Fig.~\ref{fig:060814}) is rather
similar to that of 060729, with the same final slope.  But because
060814 is somewhat fainter to begin with and has a much shorter plateau phase 
ending ten times sooner at about 7 ks, the afterglow
was only detectable by XRT for about 15~days.  With no redshift, we
estimate the jet opening angle to be $\theta_j > 10\degrees \xi$.

As our final example, we consider GRB\,061007 (Fig.~\ref{fig:061007}; \cite{Schady06}).
Unlike most XRT light curves, the X-ray light curve of 061007 has no
breaks, decaying as a single power law of slope $\alpha=1.66$ from $\sim 80$~s to $\sim
1.1$~Ms.  This object is also unusual in that the bright,
well-sampled optical light curve has exactly the same slope as the
X-ray light curve shown here.  If we interpret this burst
as being a pre-jet break decay, the opening angle is $\theta_j = 8
\degrees~n^{1/8}$.  
However, detailed consideration of the afterglow model leads to the
conclusion that it is difficult to produce a model that satisfies
the optical and X-ray observations
unless the entire observed light curve is post-jet break \cite{Schady06}.  This would require a
jet break at $t_j < 80$~s for this burst, with a resulting jet angle
$\theta_j < 1 \degrees$ \cite{Schady06}.  This solution, however,
also has difficulties, as the decay slope is shallower than expected
for a post-jet break decay of a uniform jet and does not satisfy the post-jet break
closure relations.  One solution is a very steady energy
injection phase lasting over the entire duration of the observed
afterglow, but it is hard to imagine how this can be produced.  
Alternatively, it is possible that this burst can be explained 
by a structured jet model \cite{Panaitescu05}.
GRB\,061007 raises some interesting questions about the
time-scale of the jet breaks as well as our ``standard''
interpretations of detailed X-ray and optical light curves.

\section{Cases of Perplexing Breaks}
\label{sec:perplexing}

We now consider two cases of light curve breaks that may be jet
breaks, but that do not seem completely compatible with the
expectations of the standard afterglow models.  Like GRB\,061007,
these raise interesting questions about our detailed understanding of
GRB afterglows.

   \begin{figure}
     \centering
     \parbox{2.55in}{
     \includegraphics[width=2.55in,angle=0,bb=86 31 888 620, clip]{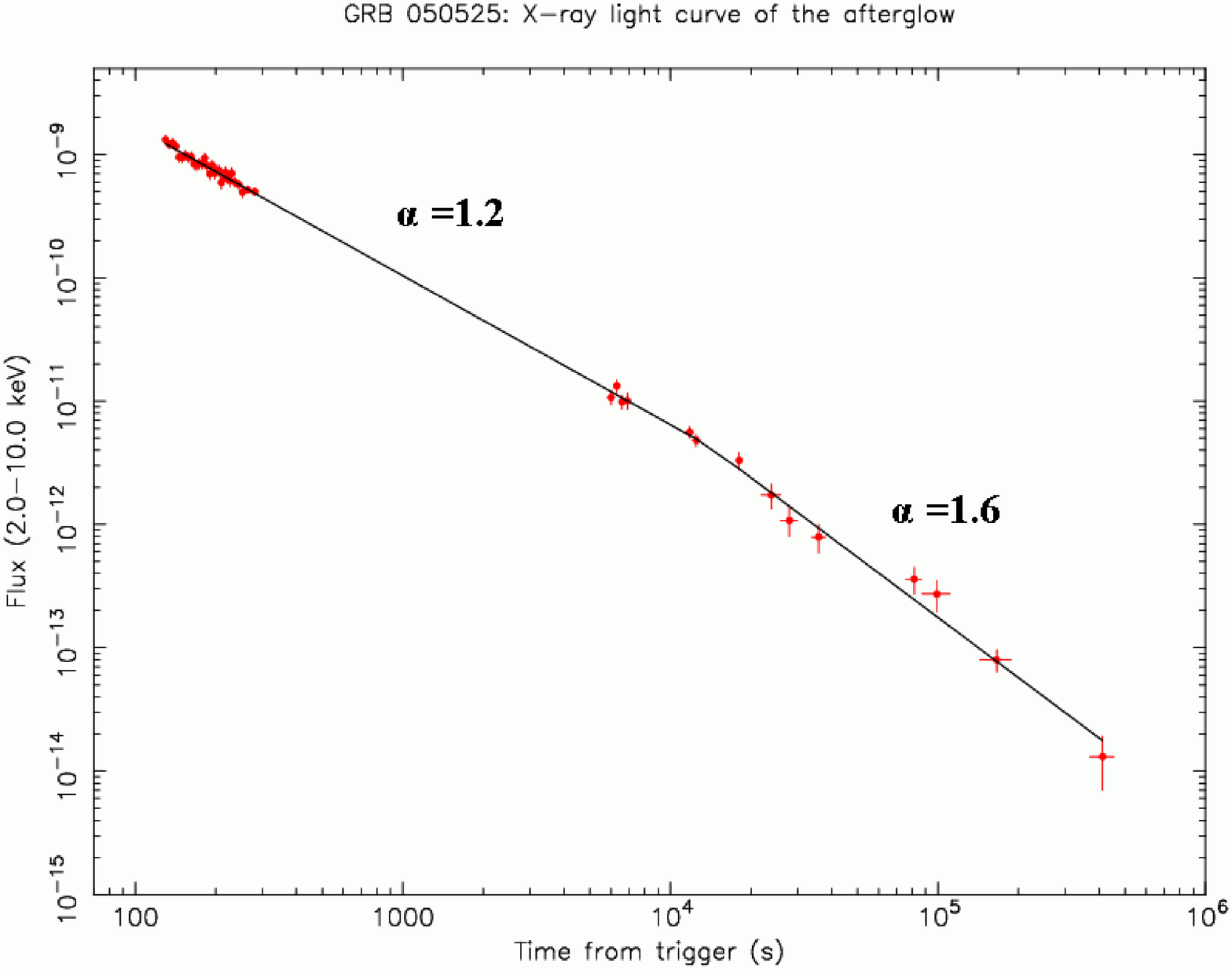}
   \caption{{\it Swift}/XRT light curve of GRB\,050525A
     \cite{Blustin06}.  There is a shallow break in the X-ray light
     curve at 13~ks to a final slope of $\alpha = 1.6$, much flatter
     than expected for a post-jet break decay.
   }
   \label{fig:050525A}}
   \hfill
     \parbox{2.55in}{
    \includegraphics[width=2.4in,angle=90,bb=20 0 550
    680,clip]{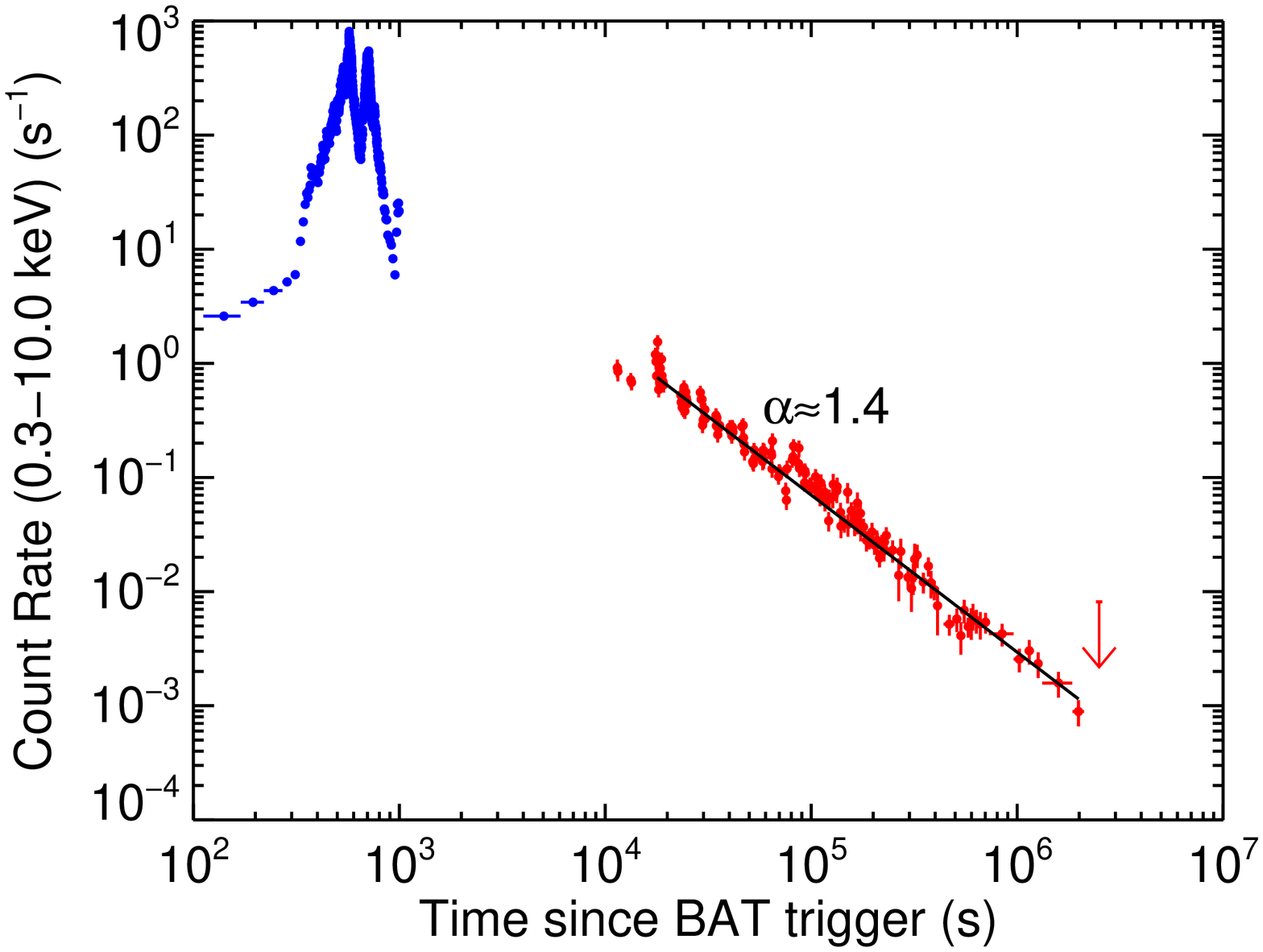}
   \vspace{-0.2in}
   \caption{{\it Swift}/XRT light curve of GRB 060124 \cite{Romano06}. The final
     break in the light curve found by \cite{Romano06} is at about 100~ks to a final slope of
     $\alpha=1.5$, though we find a single late slope fits our light curve.}
   \label{fig:060124}}
   \vspace{-0.2in}
   \end{figure}

GRB\,050525A (Fig.~\ref{fig:050525A}) is another case with good X-ray and optical light curves
\cite{Blustin06}.  The very early X-ray decay has some low-level flaring (not
shown here), followed by a two-slope decay with a shallow break from
$\alpha_1 = 1.20$ to $\alpha_2 = 1.62$, with a break time of 13.7~ks
\cite{Blustin06}.
There is a break in the optical light curve at the same time, 
making this the first achromatic break observed across both optical
and X-ray bands, and this
has therefore been interpreted as a jet break with opening angle
$\theta_j = 3\degrees$ \cite{Blustin06}.
Before this break, the data are in good agreement with the standard
fireball model for an external shock propagating into a constant density ISM.
The final decay slope is considerably flatter than expected for a jet
break, however, and does not satisfy any of the closure relations,
either pre-jet break or post-jet break.  
If this is indeed a jet break, it requires some additional component
or tweaking of the standard models (such as structured jets) to obtain a completely
self-consistent interpretation.

GRB\,060124 (Fig.~\ref{fig:060124}; \cite{Romano06}) is a fascinating
burst.  The \swift\ satellite triggered on a precursor nearly 10 minutes before the main
burst, which was therefore observed by the XRT
and UVOT instruments, producing one of the most detailed broad-band
observations of prompt emission ever made \cite{Romano06}.  Here, we
consider the long-term X-ray light curve.  \cite{Romano06} found that
the X-ray light curve of 060124 after $10^4$~s is very similar to that of
050525A, with a decay slope of $\alpha = 1.2$ from $10^4$ to $10^5$~s, followed
by a steeper decay with slope $\alpha = 1.5$.  Like 050525A, the data before the
break are in good agreement with a standard fireball model for a
constant density ISM \cite{Romano06}, but after the break the decay is
much too shallow for a standard post-jet break model.  The data may be
consistent with a structured jet model, which can accommodate a flatter
post-break decay slope \cite{Romano06}.
On the other hand, we find a
reasonably good fit to a single power law (Fig.~\ref{fig:060124}) with
small-scale flaring.

\section{Short GRBs}

Finally, we note that jet breaks have been proposed or investigated for several short
GRBs \cite{Fox05,Grupe06b,Burrows06}.  The best observations to date
are for GRBs\,050724 \cite{Grupe06b} and 051221A
\cite{Burrows06}.  Late {\it Chandra} observations of GRB\,050724 are
consistent with a slow afterglow decay with $\alpha \sim 0.9$
extending without a break until 22 days after the burst, resulting in an estimate of
$\theta_j > 25\degrees$ for an ambient density of $n=0.1~\rm{cm}^{-3}$
\cite{Berger05} and an isotropic energy of $E_{\gamma,iso}=4 \times
10^{50}$~ergs \cite{Berger05}.
By contrast, late {\it Chandra} observations of GRB\,051221A clearly
show a jet break at $t_j \sim 4$ days post-burst, breaking to a decay
slope of $\alpha \sim 1.9$.  Fitting the X-ray data, along with
optical and radio data \cite{Soderberg06}, to a detailed fireball
model, a jet opening angle of $\theta_j \sim 4\degrees - 8\degrees$ is
obtained \cite{Burrows06}.  The results suggest that short GRBs can
have a wide range of collimation angles.

\section{Discussion}

We summarize the observational details discussed above in the
following tables, which also include a few cases not discussed above.
Table 1 gives jet parameters for bursts with possible jet breaks.
These are typically given in terms of the density dependence, since we
generally do not have good estimates for densities, and assume a
radiation efficience of 0.2.  For comparison,
\cite{Frail01} assumed $n=0.1$~cm$^{-3}$, while \cite{Bloom03} use
density estimates from detailed afterglow modelling when possible
(these range from 0.1-30, but are typically on the high side of this
range), choosing a ``canonical'' value of 10~cm$^{-3}$ when no
modelling-based estimate was available.  In the case of GRB\,060428A,
which has no redshift measurement, we characterize the opening angle
in terms of $\xi$ (see Eq.~3), which we expect to be of order 1.
Table 2 gives similar parameter estimates for the bursts with no
apparent jet break.  In this case, we take the break time to be larger
than the time of the last observation, and derive corresponding
estimates of the jet opening angle and energetics.  When no redshift
is available, we again characterize the opening angle in terms of $\xi$.

\begin{table}[b]
  \caption{Jet parameters for possible jet breaks}
  \label{tab:jet breaks}
  \begin{tabular}{lccccccl}
    \hline
      !GRB     & Jet break  & Final & z & $E_{\gamma,iso,52}$ & $\theta_j$ &  $\log(E_\gamma)$ & Ref   \\
               & time (days) & $\alpha$ & & ($10^{52}$~ergs) & (deg)  & (ergs) \\
    \hline
      050315   & 2.8  & 2.1 & 1.949 & 3.3$^\dag$ & 7 $n^{1/8}$ & 50.3 & \cite{Vaughan06}\\
      050814   & 1.0  & 1.9 & 5.3   & 18  & 3 $n^{1/8}$ & 50.3 & \\
      050820A  & 14.5 & 2.2 & 2.612 & 83  & 8 $n^{1/8}$ & 51.9 & \cite{Cenko06} \\
      051221A$^a$& 4.1& 1.9 & 0.546$^b$&0.15$^b$ &  4--8 & 48.9 & \cite{Burrows06} \\
      060428A  & 9.4  & 1.8 &       &     & 9 $\xi^c$ & \\
      060614$^{a\?}$& 1.3  & 2.2 & 0.125 & 0.25& 10 $n^{1/8}$ & 49.6 & \cite{Mangano07} \\ 
    \hline
    \multicolumn{3}{l}{$^a$ Short GRB} & $^b$ from \cite{Soderberg06} & 
    \multicolumn{4}{r}{$^c$ See Eq. 3 for definition of $\xi$} \\
  \end{tabular}
  $^\dag$Over the 15-150 keV band in the
      observer's frame (from \cite{Vaughan06}), instead of the typical
      bolometric bands typically used.  This probably underestimates the jet angle.
\end{table}

\begin{table}
  \caption{Jet limits for GRBs with no jet breaks:  upper limits
    assume $t_j>t_{final}$.}
  \label{tab:no jet breaks}
  \begin{tabular}{lccccccc}
    \hline
      !GRB     & Last Break & Final & Final data  &  z & $E_{\gamma,iso,52}$ & $\theta_j$ & $\log(E_\gamma)^\dag$ \\
               & time (ks) & $\alpha$ & point (days) & &  ($10^{52}$~ergs)   & (deg)      & (ergs) \\
    \hline
      050401   & 4.9       & 1.46   & 9 & 2.9   & 35   & !$>7~n^{1/8}$ & $>51.4$ \\
      050416A  & 1.5       & 0.88   & 71 & 0.654 & 0.12 & $>42~n^{1/8}$ & $>50.5$ \\
      051109A  & $\lesssim 3$ & 1.2 & 16 & 2.345 & 5    & $>12~n^{1/8}$ & $>51.0$ \\
      060206   & $\lesssim 5$ & 1.3 & 15 & 4.048 & 5.8  & $>10~n^{1/8}$ & $>50.9$ \\
      060729   & $\sim 60$ & 1.4    & 125 & 0.54  & 1.6 & $>38~n^{1/8}$ & $>51.5$\\
      061007   & $< 0.08$  & 1.7    & 13 & 1.26  & 100  & !$>8~n^{1/8}$ & $>52.0$\\
      !!!---$^\ddag$ &     &        &    &       &      & $<0.3~n^{1/8}$&$<48.9$\\ 
    \hline
      050607   & $\sim 8$ & 1.1   & 19 &       &      & $>11~\xi$!!!  & \\
      050803   & $\sim 15$ & 1.7    & 13 &       &      & $>10~\xi$!!!  & \\
      050822   & $\sim 13$ & 1.1    & 52 &       &      & $>16~\xi$!!!  & \\
      051117A  & $\sim 15$ & 0.9    & 20 &       &      & $>11~\xi$!!!  & \\
      060202   & $\lesssim 4$ & 0.9 & 29 &       &      & $>13~\xi$!!!  & \\
      060319   & $\sim 18$ & 1.2    & 42 &       &      & $>15~\xi$!!!  & \\
      060814   & $\sim 6$  & 1.4    & 15 &       &      & $>10~\xi$!!!  & \\

    \hline
    \multicolumn{8}{l}{$^\dag$ Radiated energy assuming ambient density of 1 cm$^{-3}$}\\
    \multicolumn{8}{l}{$^\ddag$ Lower limit for 061007 jet break, assuming $t_j<80$~s}\\
  \end{tabular}
   \vspace{-0.2in}
\end{table}

The results are rather interesting.  For the few potential jet breaks,
the distribution of opening angles is similar to those of
\cite{Frail01,Bloom03} (typically $5 \degrees - 10 \degrees$), given the small sample size.
Not surprisingly, the lower limits found for the cases without breaks tend to be
considerably higher, consistent with the much larger break times
inferred for most of these cases.
Although this could be partially the result of time dilation due to
the higher redshifts in the \swift\ sample compared with earlier
studies, this cannot be the entire story.  The average redshift of our
sample is only about twice that of the earlier studies, and the ratio
of time dilation factors ($\sim 1.3$) cannot explain the much later times we find
for jet breaks.
Interpreting the early X-ray light curve breaks as jet breaks
does not seem to be a general solution to the lack of
jet break observations, since the slope following these early breaks
tends to be consistent with the standard pre-jet break afterglow models.
We are still left with the puzzle: where are the X-ray jet breaks?

We can consider whether the lack of redshifts is somehow biasing our
results.
Comparison of the jet angle limits in Table 2 for bursts with and
without redshift measurements shows that the limits are similar: the
jet limits for bursts without redshifts are typically about
$13\degrees$, very similar to the median lower limit for those bursts with
redshift measurements (though in the latter sample we have two bursts with
much higher jet angles).  We conclude that there is no evidence that
the jet angle limits for bursts without redshifts differ from those
with redshifts.

The lower limits to jet angles for bursts without observed jet breaks are
typically about a factor of two higher than the jet angles found for
bursts with jet breaks around a few days.
If these populations are the same, we can reconcile these results by
assuming a smaller density for the bursts without breaks.  Because the dependence of
$\theta_j$ on $n$ is so weak, very low densities are required, with
$n\sim 4 \times 10^{-3}$~cm$^{-3}$, typical of densities expected in a hot phase
of the ISM.
Alternatively, the radiation  efficiency could be much lower than the
value of 0.2 assumed in Table 2: this would require $\eta_\gamma \sim
10^{-3}$ for these bursts.

Within the context of the standard jet break model (for a uniform
jet), we therefore must consider the possibility of a bimodel
distribution of either density or efficiency.  We note that a bimodal
density distribution might in principle be able to explain why bursts
with bright optical afterglows seem to have earlier jet break times than
those found in the X-ray sample: since the cooling frequency is
typically between the X-ray and optical bands, the low frequency
flux should be density-dependent while the X-ray flux is not.  The
observations may imply a bimodal ambient density,
with low densities for some GRBs (which tend to have very
late jet breaks and faint optical counterparts). 

The derived beaming-corrected energies for the bursts with breaks are
comparable to those found in earlier studies.
However, the lower limits on $E_\gamma$ for the bursts without
breaks are comparable to the average energies found by \cite{Bloom03}. 
Unless the jet breaks are typically happening just after the end of
our observations, the concept of a constant
energy reservoir \cite{Frail01} may need to be revisited.  Deep
observations at very late times are needed to make further progress, and
we have an approved {\it Chandra} ToO program to begin to address this.

Finally, we reiterate the point that many X-ray afterglows seem to
exhibit characteristics that are difficult to accommodate in the
standard fireball model.  We have shown a number of cases in which the
decay cannot be directly reconciled with the expected closure
relations.  Some of these can be explained by extended energy
injection phases, but the mechanism needs to be quite steady, and in
the (admittedly extreme) case of GRB\,061007, needs to extend continuously
for over six days in the GRB rest frame.  
A more plausible solution may lie in structured jets, which could also
explain some other cases in which apparent jet breaks are followed by
rather shallow decays.
On the other hand, similar problems with the
standard model have been raised by unexpected chromatic breaks
\cite{Panaitescu06} at early times and by the lack of an X-ray break
accompanying the optical break in GRB\,060206.
Many of these features can be explained by tweaking the knobs of the
fireball model, but one begins to have an uncomfortable feeling that
we may be missing something more fundamental that these data are
trying to tell us.  Perhaps these observations are the clues that will
ultimately lead to a better understanding of the underlying physics
behind GRB afterglows and the formation of jets during black hole formation.

\acknowledgments
This work is supported by NASA contract NAS5--00136, and would not have
been possible without the contributions of our colleagues at Penn State
University, the University of Leicester, the INAF-Osservatorio Astronomico
di Brera, INAF-IASFPA, the ASI Science Data Center, and NASA/Goddard
Space Flight Center.  We would like to particularly thank Michael Stroh, Loredana
Vetere, and Vanessa Mangano for help in preparing the light curves, and Bing
Zhang and Dick Willingale for useful discussions of this topic.


\end{document}